\documentclass[11pt]{report}
\usepackage{geometry}                
\usepackage{graphicx}
\usepackage{amssymb}
\usepackage{epstopdf}
\begin{document}
\begin{large}
\begin{center}
  {\Large \bf Letter of intent; Precise measurements of very forward particle production at RHIC}
  \vskip 10mm
  \vskip 20mm

{\large Y.Itow, H.Menjo, G.Mitsuka, T.Sako} \vskip 3mm  
{\normalsize Solar-Terrestrial Environment Laboratoy / Kobayashi-Maskawa Institute for the Origin of Particles and the Universe / Graduate School of Science, Nagoya University, Japan} \vskip 10mm

{\large K.Kasahara, T.Suzuki, S.Torii}  \vskip 3mm 
{\normalsize Waseda University, Japan}  \vskip 10mm

{\large O.Adriani, A.Tricomi}  \vskip 3mm
{\normalsize INFN, Italy}  \vskip 10mm

{\large Y.Goto}  \vskip 3mm
{\normalsize Riken BNL, Japan}  \vskip 10mm

{\large K.Tanida}  \vskip 3mm
{\normalsize Seoul National University}  \vskip 10mm

{\large}  \vskip 3mm
{\normalsize}  \vskip 10mm

{\large}  \vskip 3mm
{\normalsize}  \vskip 10mm
\end{center}

\clearpage
\begin{center}
  {\bf ABSTRACT}
\end{center}
In this letter of intent, we propose an experiment for the precise measurements of very forward particle production 
at RHIC.
The proposal is to install a LHCf-like calorimeter in the ZDC installation slot at one of the RHIC interaction
points.
By  installing a high-resolution electromagnetic calorimeter at this location we measure
the spectra of photons, neutrons and $\pi^{0}$ at pseudorapidity $\eta>$6.

The new measurements at 500\,GeV p-p collisions contribute to improve the hadronic interaction models
used in the cosmic-ray air shower simulations.
Using a similar kinematic coverage at RHIC to that of the measurements at LHC 7--14\,TeV p-p collisions,
we can test the Feynman scaling with a wide $\sqrt{s}$ range and make the extrapolation of models
into cosmic-ray energy more reliable.
Combination of a high position resolution of the LHCf detector and a high energy resolution of the ZDC 
makes it possible to determine p$_{T}$ of forward neutrons with the ever best resolution.
This enables us to study the forward neutron spin asymmetry discovered at RHIC in more detail. 

Another new experiment expected at RHIC is world-first light-ion collisions.
Cosmic-ray interaction models have been so far tested with accelerator data, but colliders have provided
only p-p (or \={p}) and heavy-ion collisions. 
To simulate the interaction between cosmic-ray particles and atmosphere, collision of light ions like nitrogen 
is a ultimate goal for the cosmic-ray physics.
We propose 200\,GeV p-N collisions together with 200\,GeV p-p collisions to study the nuclear effects in the 
forward particle production. 

The experiment can be performed by using the existing LHCf detector.
Considering the geometry and response of one of the LHCf detectors, we propose some short dedicated
operations.
For the particle spectrum and nuclear effect measurements, we need several hours of operation at each
condition with ideal luminosities  6$\times$10$^{29}$\,cm$^{-2}$~s$^{-1}$, 
4$\times$10$^{31}$\,cm$^{-2}$~s$^{-1}$ and 7$\times$10$^{30}$\,cm$^{-2}$~s$^{-1}$ at 500\,GeV p-p, 
200\,GeV p-p and 200\,GeV p-N collisions, respectively. 
We require collisions of unsqueezed beams to reduce the angular divergence at collisions.
For the spin asymmetry measurement, we require 11 hours of operation at 500\,GeV p-p collisions
with a luminosity 2.5$\times$10$^{30}$\,cm$^{-2}$~s$^{-1}$.
Horizontal spin polarization is also required.

Our basic idea is to bring one of the LHCf detectors to RHIC after the LHC 13\,TeV p-p collision runs 
planned  in early 2015, and then operate from 2016 season at RHIC.

\clearpage
\tableofcontents
\clearpage

\chapter{Introduction} \label{sec-introduction}
{\bf Improving cosmic-ray air shower simulation}\\
The origin of cosmic rays is a century-standing problem.
Recent observations of ultra-high-energy cosmic rays (UHECR) by the Pierre Auger Observatory (PAO) \cite{PAO}
and Telescope Array \cite{TA} have been dramatically improved in both the statistics and the systematics.
The existence of a spectral cutoff at approximately 10$^{19.5}$\,eV is now clearly identified.
However, the interpretation of the observed results is not settled.
One of the main reasons for the difficulty is the uncertainty in air shower modeling.
Fig.\ref{fig-PAO-composition} shows the so-called X$_{max}$ parameter as a function of the cosmic-ray energy
observed by the PAO.
Here, X$_{max}$ is the height of the shower maximum measured from the top of the atmosphere in g/cm$^{2}$.
Experimental data are compared with the predictions by air shower simulation with the two extreme assumptions
that all cosmic rays are protons or iron nuclei.
Four lines in each assumption are due to the use of different interaction models in the air shower simulation.
Because of the fact that the difference between models is larger than the experimental errors,
the determination of the primary chemical composition is difficult, and hence, the nature
of the spectral cutoff is not concluded.
The determination of the chemical composition at 10$^{17}$\,eV is also important because, at approximately
this energy, the source of the cosmic rays is believed to switch from galactic to extragalactic and the chemical 
composition rapidly changes with energy \cite{KASCADE}.
However, due to the uncertainty in air shower modeling, the determination of the chemical composition at this
energy range is still model dependent. 
To solve the origin of mysterious UHECRs and to confirm the standard scenario of the
cosmic-ray origin, constraints from the accelerator experiments are indispensable.

  \begin{figure} 
  \begin{center}
  \includegraphics[width=10cm]{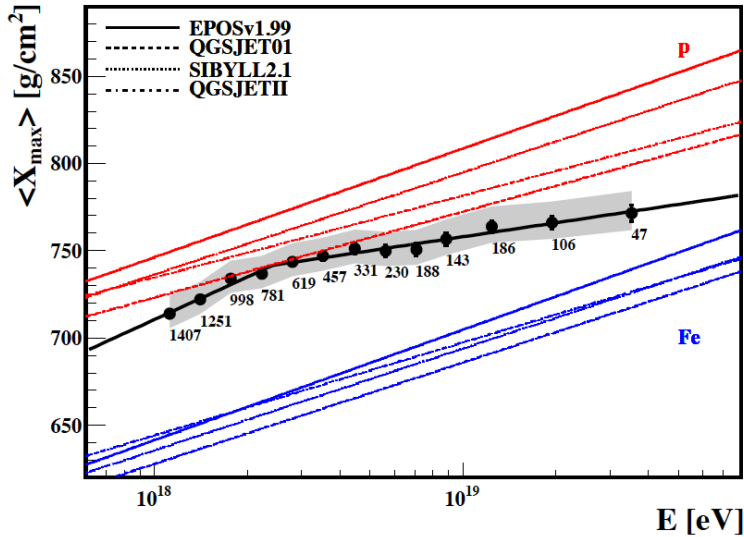} 
  \vspace*{8pt}
  \caption{X$_{max}$ of air showers observed by the Pierre Auger Observatory. \cite{PAO}
  \label{fig-PAO-composition}}
  \end{center}
  \end{figure}

The difficulty in modeling hadronic interactions, which is essential to determine the air shower development,
is due to the difficulty in modeling the soft interaction described by non-perturbative QCD.
Experimentally, particles produced in such processes have a large energy flux in the forward direction and are 
difficult to measure especially in the collider experiments.
Cosmic-ray interaction models have been tested with a variety of accelerator experiments with a limited
number of forward measurements, and most of the data thus far are limited in proton-proton (or anti-proton) collisions.
The Large Hadron Collider (LHC) provides an unprecedented quality of test data; this facility gives the highest
collision energy, and the experiments cover a very wide range of pseudorapidity ($\eta$) \cite{dEntteria}.
The same quality of lower-energy collision data is very important to test the $\sqrt{s}$ dependence of hadronic 
interactions to extrapolate the models into the UHECR energy region.
In contrast to p-p collisions, only d-Au collisions at RHIC and  p-Pb collisions at LHC provide collision situations
that are similar to cosmic-ray protons interacting with the atmosphere.
In both cases, strong nuclear effects were reported by STAR \cite{STAR} and ALICE \cite{ALICE}.
These effects will be important inputs to simulate proton-atmosphere collisions at extreme conditions.
However, no direct tests of nuclear effects in proton-atmosphere collisions have been performed thus far.

Large Hadron Collider forward (LHCf) is one of the LHC experiments to measure forward neutral particles to 
calibrate the interaction models used in the cosmic-ray physics \cite{LHCf}.
LHCf successfully acquired to take data for LHC 900\,GeV, 2.76\,TeV and 7\,TeV
proton-proton collisions and 5.0\,TeV ($\sqrt{s_{NN}}$) p-Pb collisions.
LHCf installed compact calorimeters at the installation slot of the Zero Degree Calorimeter (ZDC) located 140\,m
from an interaction point of the LHC.
Two independent detectors called Arm1 and Arm2 at either side of the interaction point were installed.
At this location, neutral particles (predominantly photons decayed from $\pi^{0}$ and neutrons) emitted at
$\eta>$8.4 are observed.
Each detector has two small calorimeter towers that allow simultaneous detection of two high-energy particles
and hence identification of photon pairs originating from $\pi^{0}$ by reconstructing the invariant mass of these
particles.
The calorimeters are optimized to measure TeV photons and have energy and position resolutions
of $8\%/\sqrt{E/100\,GeV}+1\%$ and $<$200\,$\mu$m, respectively, in the LHC environment.
These dedicated electromagnetic calorimeters enabled the first high-resolution measurements of 
electromagnetic showers at approximately zero degrees at colliders.
These measurements are motivated by the desire to constrain the source of the mesonic branch in an air shower, 
but it is also possible to access the baryonic part by measuring forward neutrons with limited resolutions.
Thus far, LHCf has published energy spectra of forward photons at 900\,GeV \cite{LHCf900GeV} and 7\,TeV \cite{LHCf7TeV} 
and forward $\pi^{0}$ spectra at 7\,TeV \cite{LHCfpi0} as shown in Fig.\ref{fig-lhcf-photon} and Fig.\ref{fig-lhcf-pi0}.
Generally, no interaction model perfectly explains the LHCf results, but the models bracket the experimental data well.
As mentioned above, comparisons of data at different collision energies are essentially important,
but a small p$_{T}$ coverage in the 900\,GeV collision data does not allow an effective comparison
with the 7\,TeV data.

  \begin{figure} 
  \begin{center}
  \includegraphics[width=13cm]{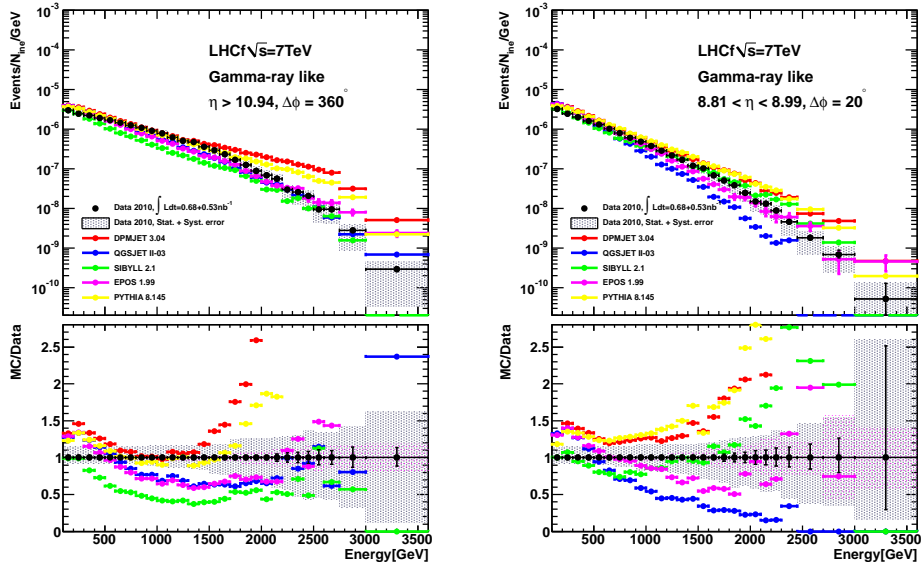} 
  \vspace*{8pt}
  \caption{Forward photon spectra measured by LHCf at the LHC 7\,TeV p-p collisions.
  \label{fig-lhcf-photon}}
  \end{center}
  \end{figure}

  \begin{figure} 
  \begin{center}
  \includegraphics[width=10cm]{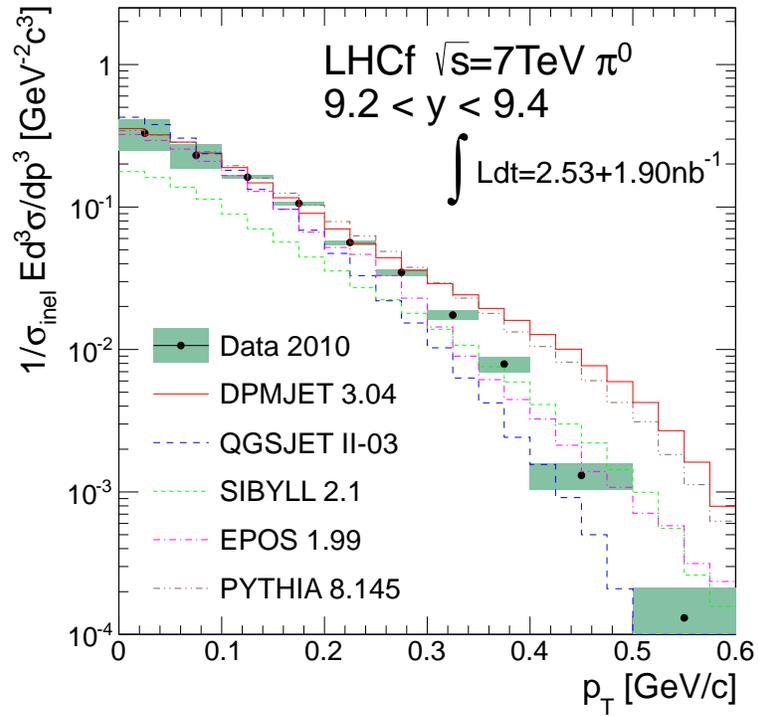} 
  \vspace*{8pt}
  \caption{An example of forward $\pi^{0}$ spectra measured by LHCf at the LHC 7\,TeV p-p collisions. 
  \label{fig-lhcf-pi0}}
  \end{center}
  \end{figure}

\vskip 5mm
\noindent {\bf Extended measurements of forward neutron spin asymmetry}\\
With the first polarized p-p collisions at $\sqrt{s}$ = 200\,GeV at RHIC, 
a large single transverse-spin asymmetry ($A_N$) for neutron production 
in very forward kinematics was discovered by a polarimeter development 
experiment \cite{Fukao:2006vd}.
The discovery of the large $A_N$ for neutron production is new, 
important information to understand the production mechanism of the very 
forward neutron. 
 
  \begin{figure}[htbp]
  \begin{center}
  \includegraphics[width=7.2cm]{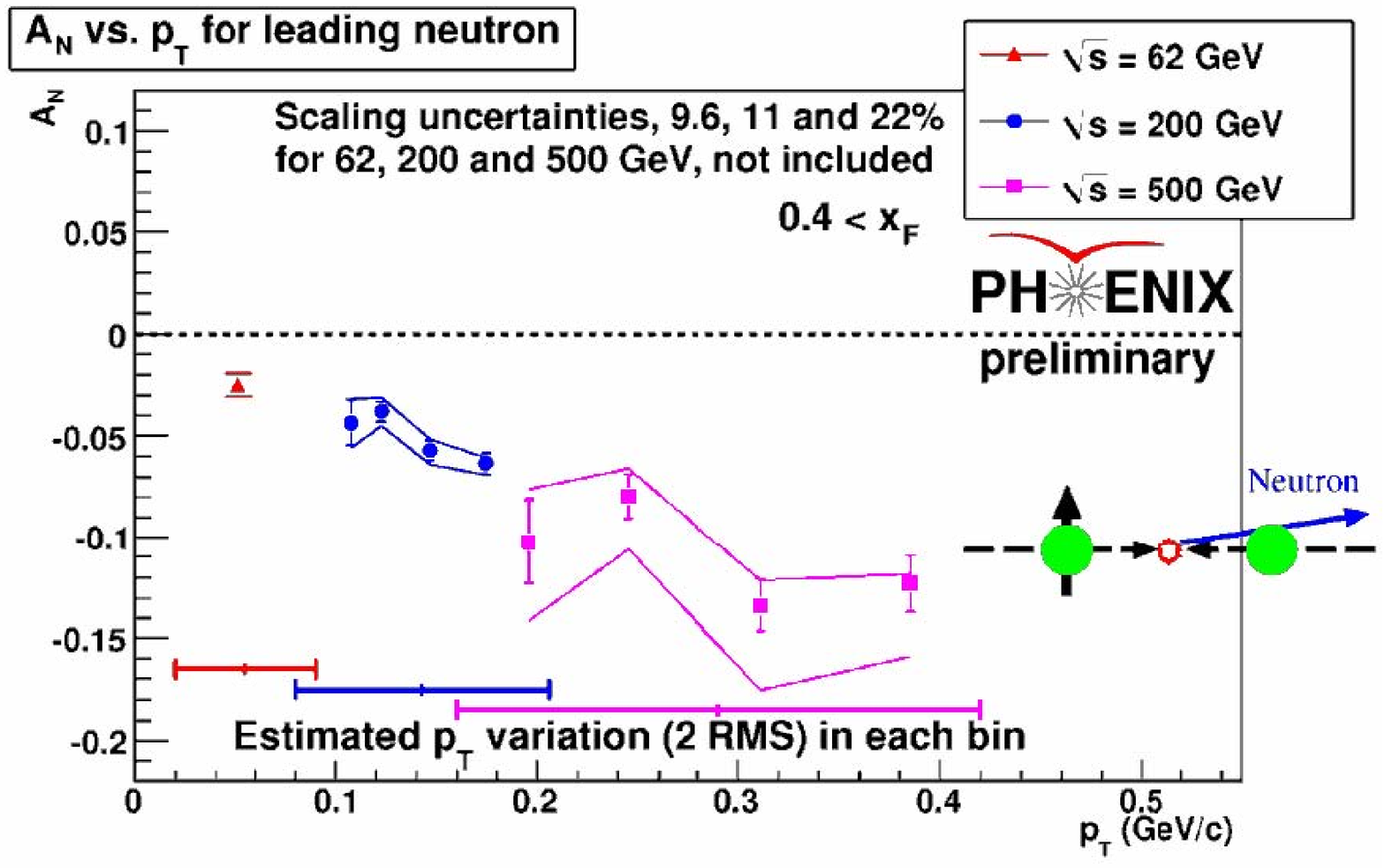}
  \includegraphics[width=7.2cm]{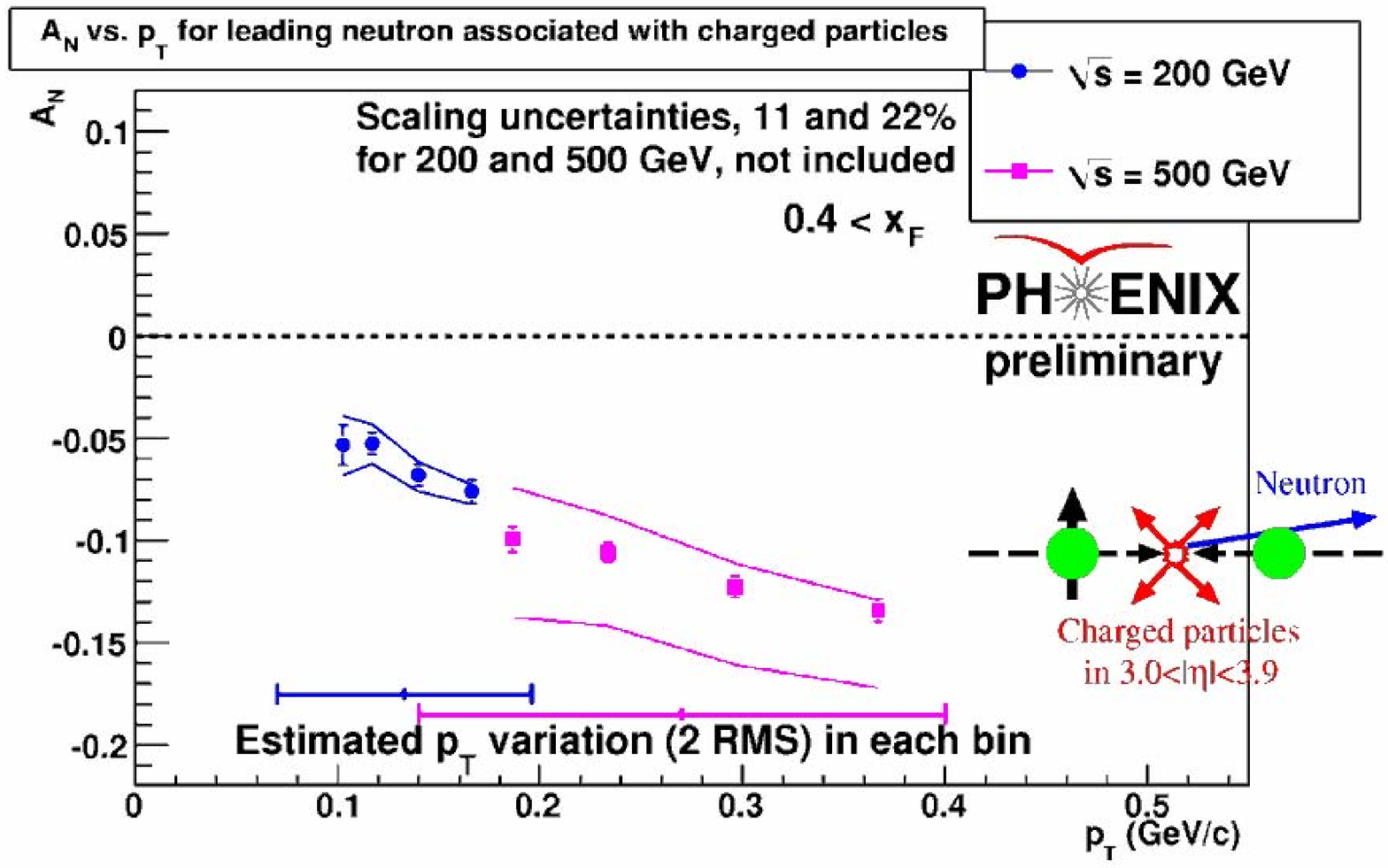}
  \caption{The measured asymmetries of very forward neutron production as 
  functions of $p_T$ with an inclusive-neutron trigger (left) and with 
  a semi-inclusive neutron trigger including a beam-beam interaction 
  requirement (right).}
  \label{fig:asym_pt}
  \end{center} 
  \end{figure}

The $\sqrt{s}$ dependence of $A_N$ from three different collision energies, 62.4\,GeV, 200\,GeV, and 
500\,GeV was studied. 
The result is shown in Fig.~\ref{fig:asym_pt}. 
The hit position dependence on the detector was measured at each energy, 
although this dependence was largely smeared by the position resolution. 
The result was converted to the $p_T$ dependence, which showed a hint of the $p_T$ scaling property 
of $A_N$ of the very forward neutron production. 
The asymmetry is caused by interference between spin-flip and non-flip 
amplitudes with a relative phase. 
Kopeliovich et al. \cite{Kopeliovich:2011bx} studied the interference of a pion and $a_1$, or a pion and $\rho$ 
in the $1^+S$ state. 
The data agreed well with the independence of energy. 
The asymmetry is sensitive to the presence of different mechanisms, e.g., Reggeon exchange with 
spin-non-flip amplitudes even if these amplitudes are small. 

\vskip 5mm
In this letter of intent, we propose an experiment like LHCf to be performed at the Relativistic Heavy Ion 
Collider at BNL to improve modeling of air shower simulations and understanding of the nature of the forward
neutron spin asymmetry.
By installing a high-resolution electromagnetic calorimeter at the ZDC installation slot of RHIC, particle production
at approximately zero degrees is studied with unprecedented resolutions. 
The proposed experiment is temporally called RHICf in this letter.  

\chapter{Physics} \label{sec-physics}
At RHIC, the ZDC installation slot is located at 18\,m from the IP.
The 10\,cm gap between beam pipes allows the installation of a LHCf detector itself and measurements of 
forward neutral particles down to $\eta$=6.
Here, we discuss physics of neutral particle measurements at $\eta>$6 at the RHIC energy.
Three advantages, two related to cosmic-ray physics and one for spin asymmetry, are introduced.
the details of the detector configuration are discussed in Chap.\ref{sec-setup}.

\section{$\sqrt{s}$ scaling of hadronic interaction}
Fig.\ref{fig-pi0-edependence} shows the energy spectra of all $\pi^{0}$ at $\sqrt{s}$= 500\,GeV, 7\,TeV and 
50\,TeV (E$_{lab} = $1.3$\times$10$^{18}$\,eV) predicted by the DPMJET3 and QGSJET-II \cite{QGS} models.
With the particle energy scaled by the beam energy ($\sim$Feynman X, $x_{F}$), the DPMJET3 model assumes 
a perfect Feynman scaling \cite{Feynman}, whereas QGSJET-II shows a softening in higher energy collisions.
The assumption of the Feynman scaling or collision energy dependence is an important issue to be tested at 
the collider experiments.
Fig.\ref{fig-phasespace} shows the photon spectra, $d^{2}N/dp_{T}/dE$, in the energy-p$_{T}$ plane 
predicted by the DPMJET3 interaction model \cite{DPM} for $\sqrt{s}$= 500\,GeV, 900\,GeV and 7\,TeV 
proton-proton collisions.
The red triangles indicate the phase spaces $\eta>$6.0 for 500\,GeV and $\eta>$8.4 for the other energies
corresponding to the RHICf and LHCf cases, respectively.
The phase-space coverage is very similar between 500\,GeV at RHIC and 7\,TeV at LHC, but is very narrow 
in the LHC 900\,GeV case.
The wide and similar coverage of RHIC 500\,GeV p-p collisions and LHC 7-14\,TeV p-p collisions provides a strong
constraint for the Feynman scaling hypothesis that is important to extrapolate the interaction models into the 
UHECR region.  

   \begin{figure}
  \begin{center}
  \includegraphics[width=8cm]{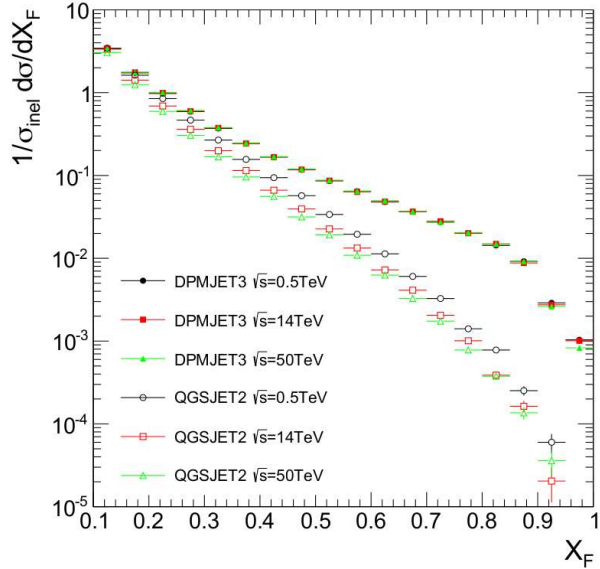}
  \vspace*{8pt}
  \caption{X$_{F}$ spectra of all $\pi^{0}$ at $\sqrt{s}$ = 500\,GeV, 7\,TeV and 50\,TeV p-p collisions.
  \label{fig-pi0-edependence}}
  \end{center}
  \end{figure}

  \begin{figure}
  \begin{center}
  \includegraphics[width=6.5cm, height=6cm]{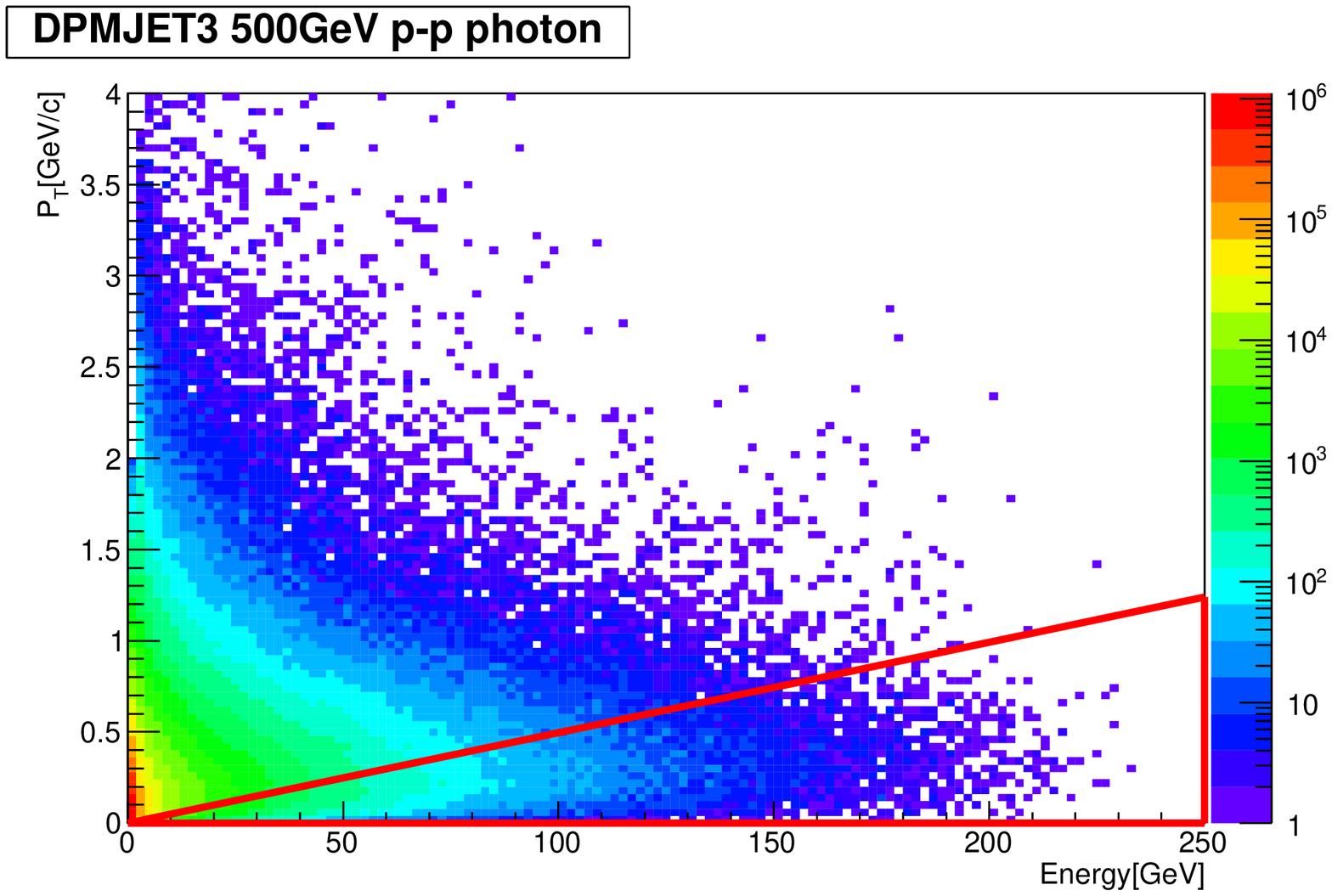}\\
  \includegraphics[width=6.0cm]{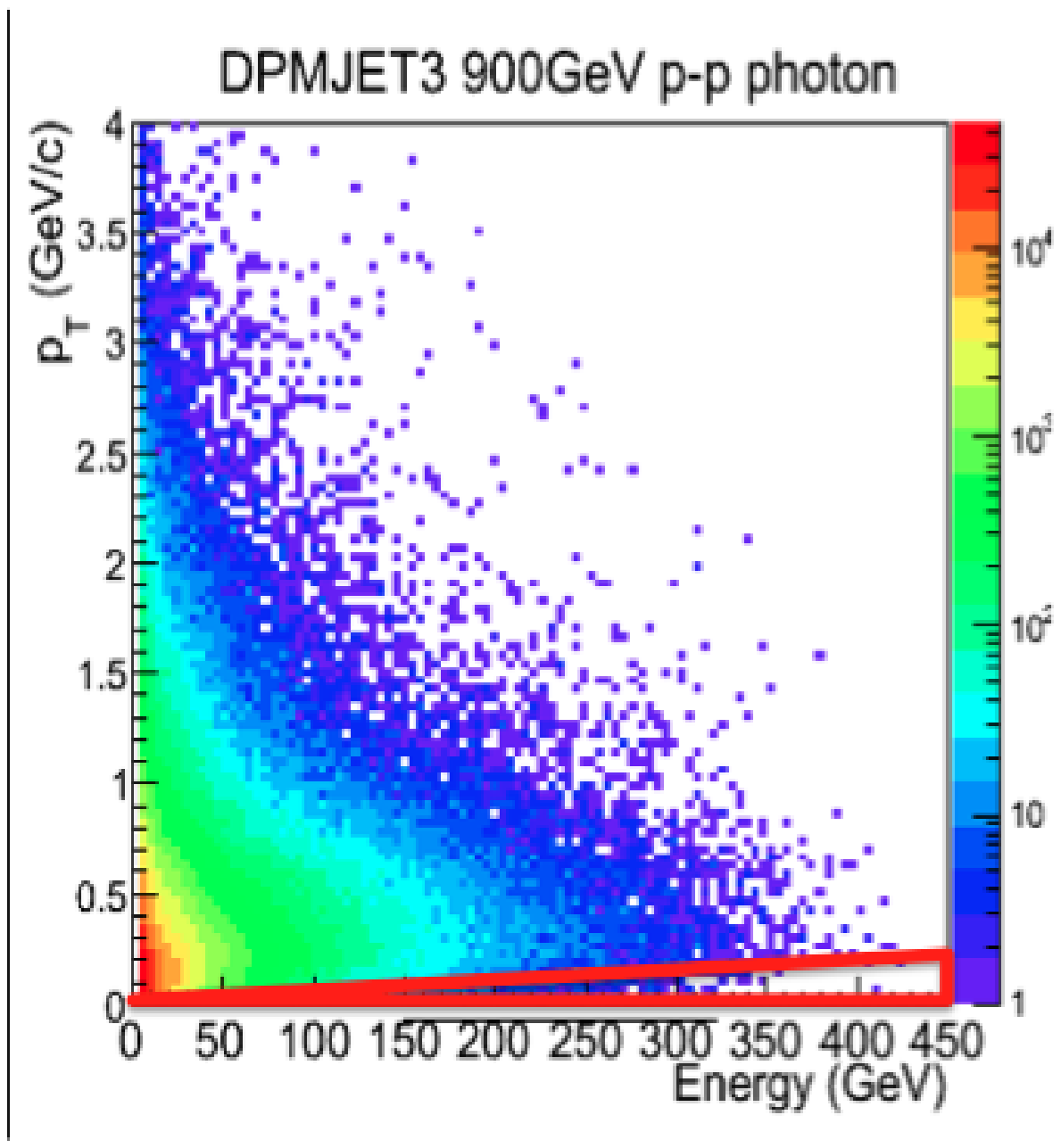}
  \includegraphics[width=6.7cm]{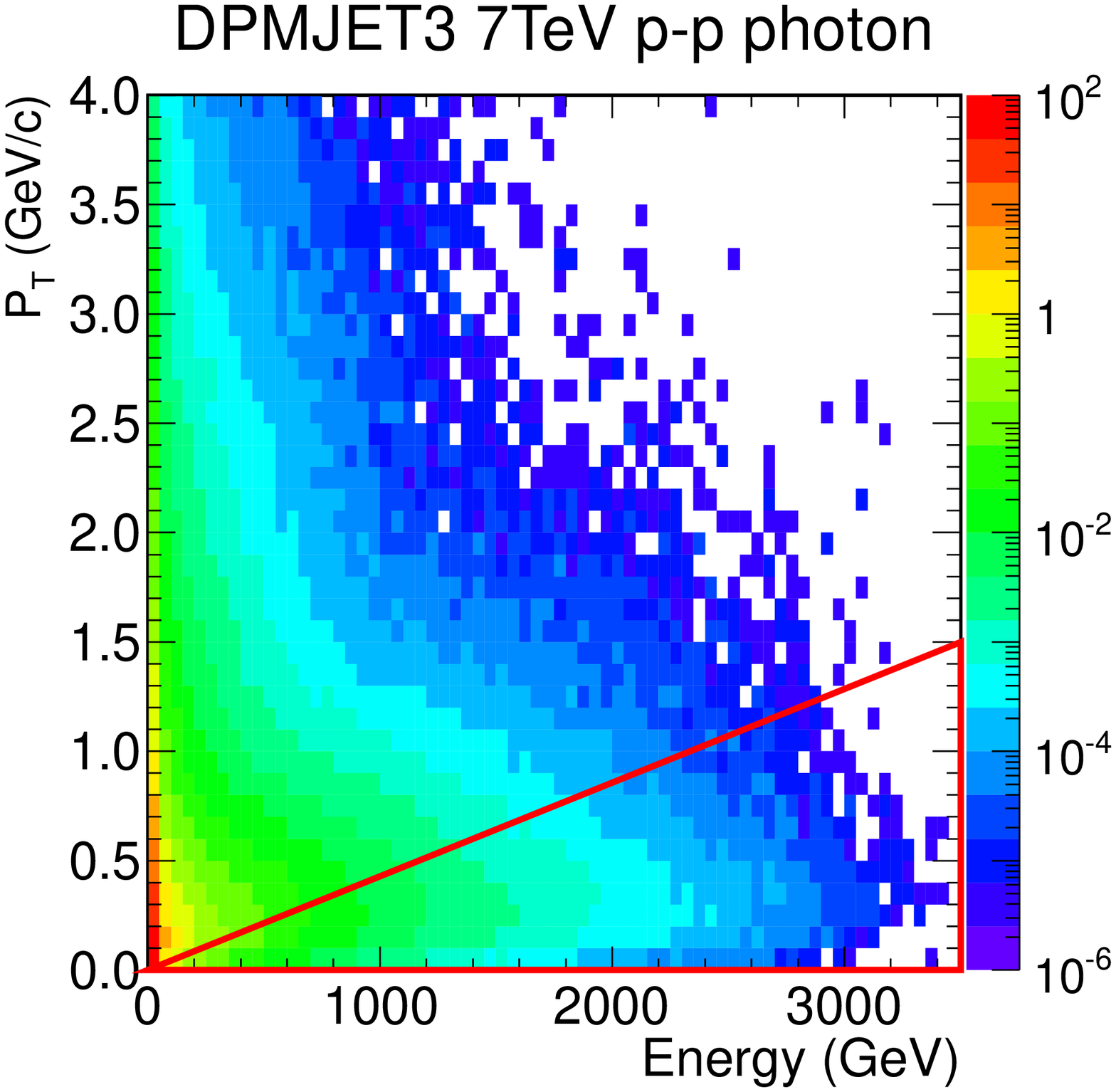}
  \vspace*{8pt}
  \caption{Photon yield in the energy-p$_{T}$ phase space at $\sqrt{s}$=500\.GeV, 900\,GeV and 7\,TeV p-p collisions
  predicted by the DPMJET3 model.   Red triangles indicate the geometrical acceptances of a LHCf-like experiment 
  installed at RHIC in 500\,GeV and at LHC in 900\,GeV and 7\,TeV collisions.
  \label{fig-phasespace}}
  \end{center}
  \end{figure}

The configuration at RHIC allows the identification of $\pi^{0}$ using the double towers of the LHCf detector.
The minimum opening angle of a photon pair decayed from a $\pi^{0}$ of energy E is described as
$\theta_{min}$ = 2M/E, where M is the rest mass of $\pi^{0}$.
In case of 900\,GeV p-p collisions at LHC, $\theta_{min}$=600\,$\mu$rad is close to the maximum aperture
of the LHCf detector (90\,mm/140\,m) and no photon pair is identified.
However, in the 500\,GeV p-p collision at RHIC,  $\theta_{min}\sim$1\,mrad is sufficiently smaller than
the RHICf aperture (90\,mm/18\,m = 5\,mrad) to allow the detection of photon pairs.
Very forward $\pi^{0}$ production was measured by the UA7 experiment for Sp\={p}S 630\,GeV proton-antiproton
collisions \cite{UA7}.
Because UA7 covered slightly higher p$_{T}$ than the RHICf configuration at similar collision energy, 
RHICf can provide complementary information to understand the meson production spectrum. 

\section{Nuclear effect in atmospheric nuclei} \label{sec-phys-nuc}
Direct collider measurements of light ions are expected for cosmic-ray physics,
and RHIC is expected to realize the first light-ion collision.
Because the nuclear effect is not explicitly implemented in the models, some steps are necessary to clarify the
model dependence of the nuclear effect. 
Fig.\ref{fig-mod-factor}  shows energy spectra of $\pi^{0}$ and neutrons at $\eta>$6 of $\sqrt{s_{NN}}$=200\,GeV
p-N collisions using three interaction models, DPMJET3, QGSJET II and EPOS \cite{EPOS}.
Here `N' designates nitrogen.
As references, spectra in the p-p collisions using the same models in the same rapidity range are also shown.
The ratio, R$_{model}$, of the spectra in p-N collisions to those in p-p collisions are shown in Fig.\ref{fig-nuc-effect}.
These ratios are thought to represent the nuclear effect implemented in each model.
To enhance the difference between models, double-ratio spectra R$_{QGSJET II}$/R$_{DPMJET3}$ and
R$_{EPOS}$/R$_{DPMJET3}$ are shown in Fig.\ref{fig-nuc-double}.
The results indicate that QGSJET II and EPOS assume similar nuclear effects in the $\pi^{0}$ production
spectrum but are different in amplitude at approximately a 20\% level.
However, QGSJET II and EPOS predict harder and softer neutron spectra, respectively, than DPMJET3,
with a 40\% difference.
The nuclear effect of the very forward particle production in light-ion collisions is a very new field and is 
never tested in the collider experiments. 
Experimental verification of different model predictions becomes possible for the first time at RHIC.

  \begin{figure}
  \begin{center}
  \includegraphics[width=7cm]{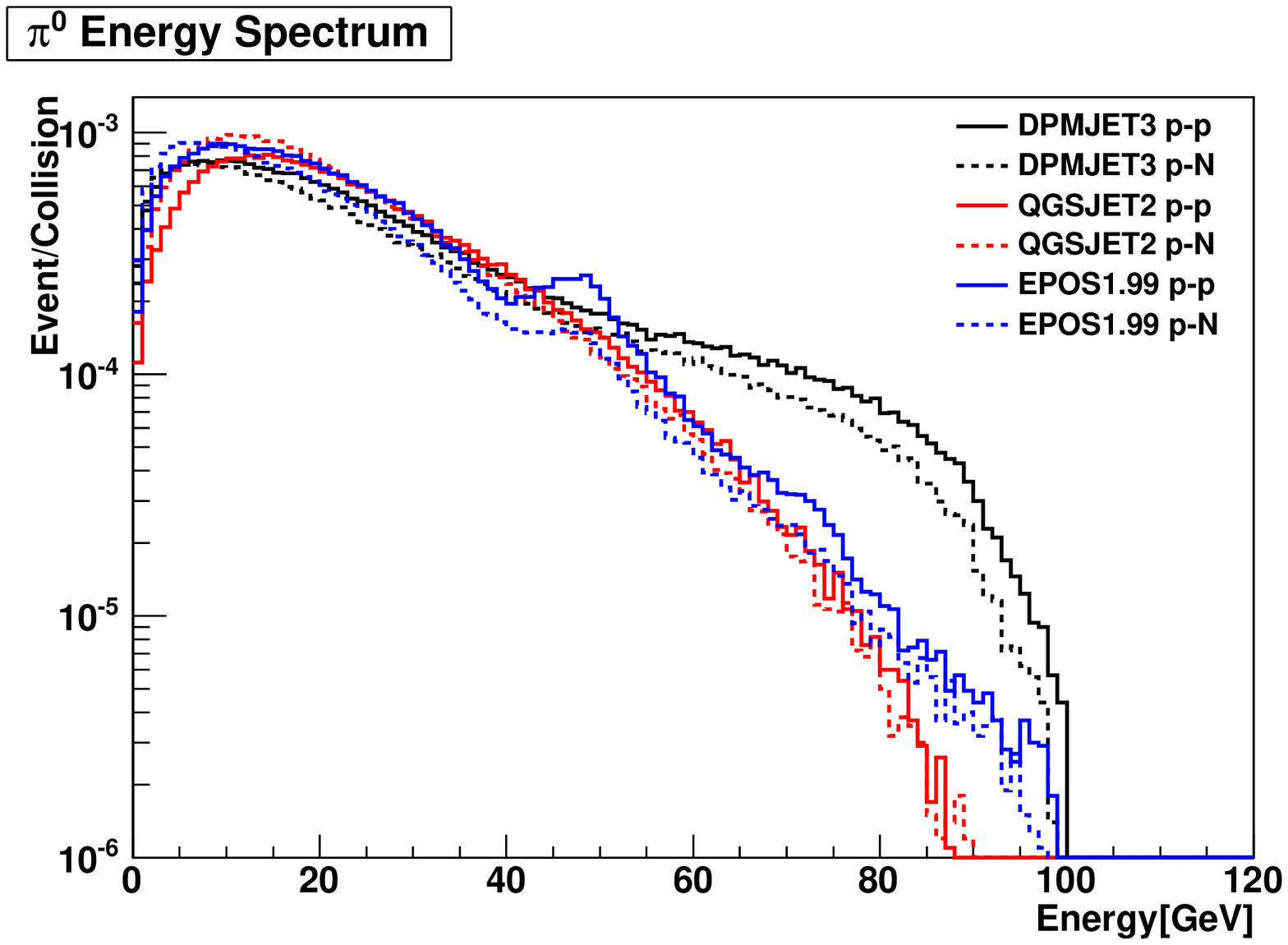}
  \includegraphics[width=7cm]{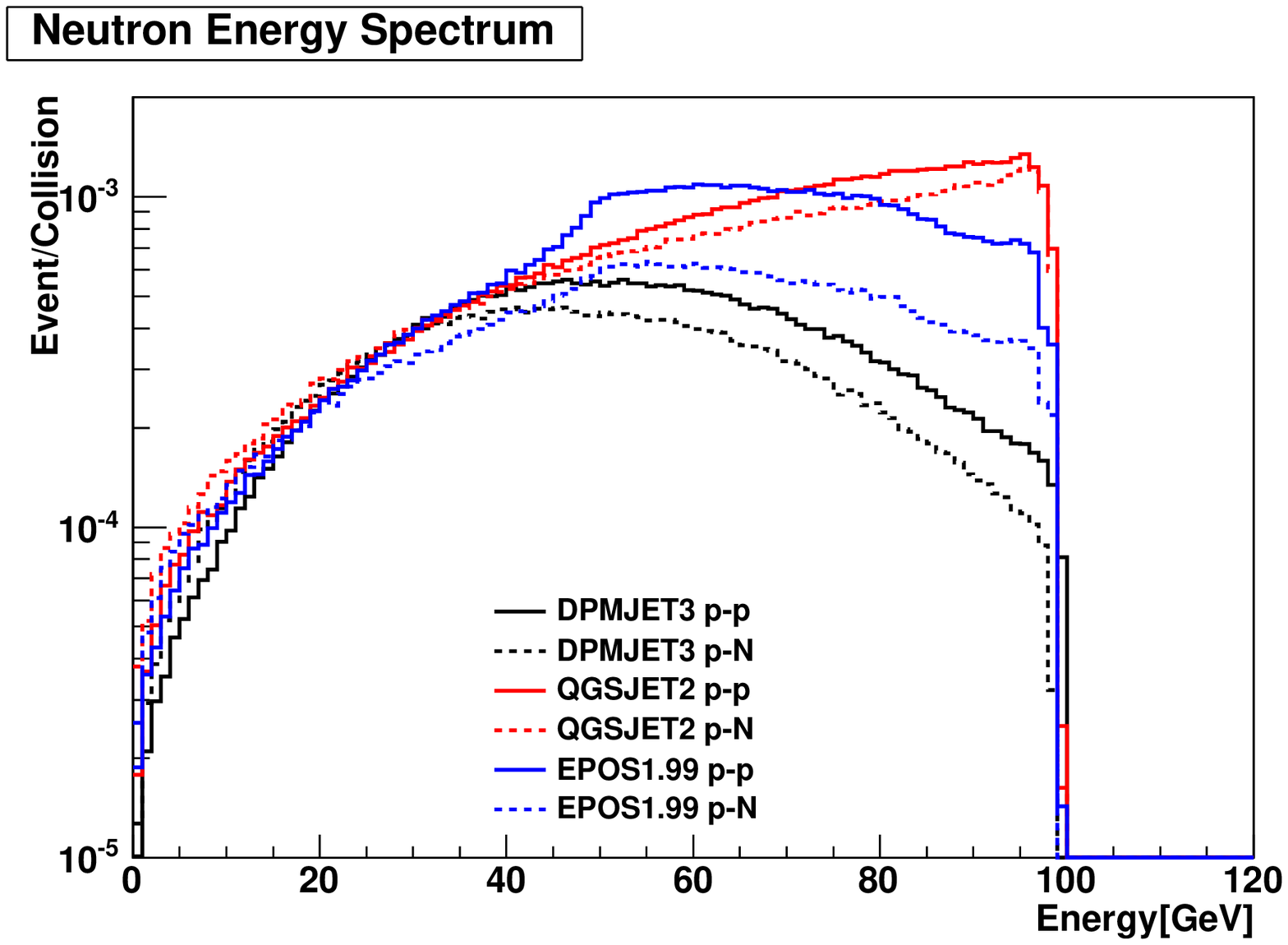}
  \vspace*{8pt}
  \caption{Energy spectra of photons (left) and neutrons (right) found at $\eta>$6 of the proton remnant side in 
  $\sqrt{s_{NN}}$=200\,GeV p-nitrogen collisions predicted by the DPMJET3, QGSJET-II and EPOS models.
  As references, the same spectra in the p-p collisions are also indicated by dashed lines. 
  \label{fig-mod-factor}}
  \end{center}
  \end{figure}

  \begin{figure}
  \begin{center}
  \includegraphics[width=7cm]{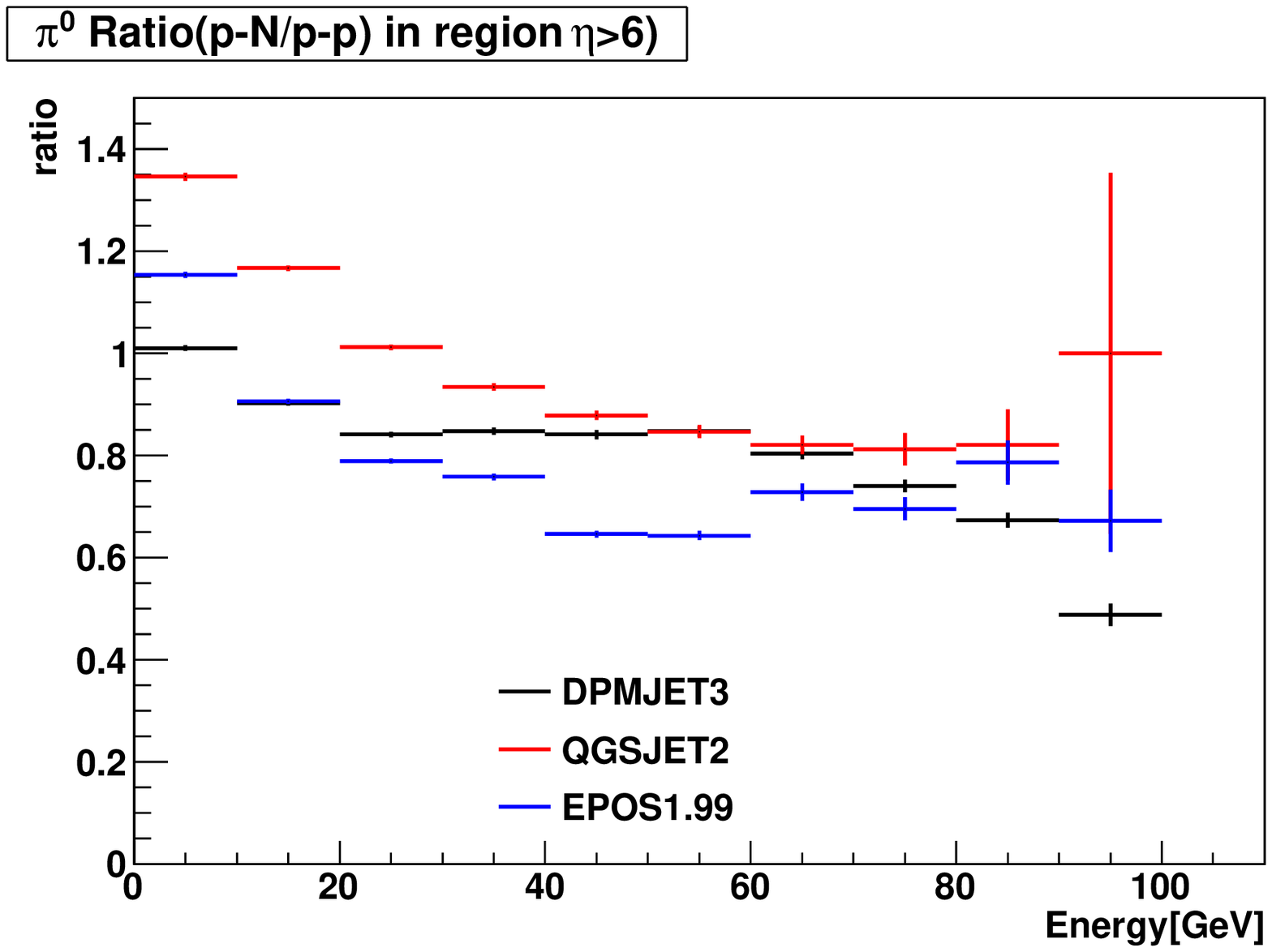}
  \includegraphics[width=7cm]{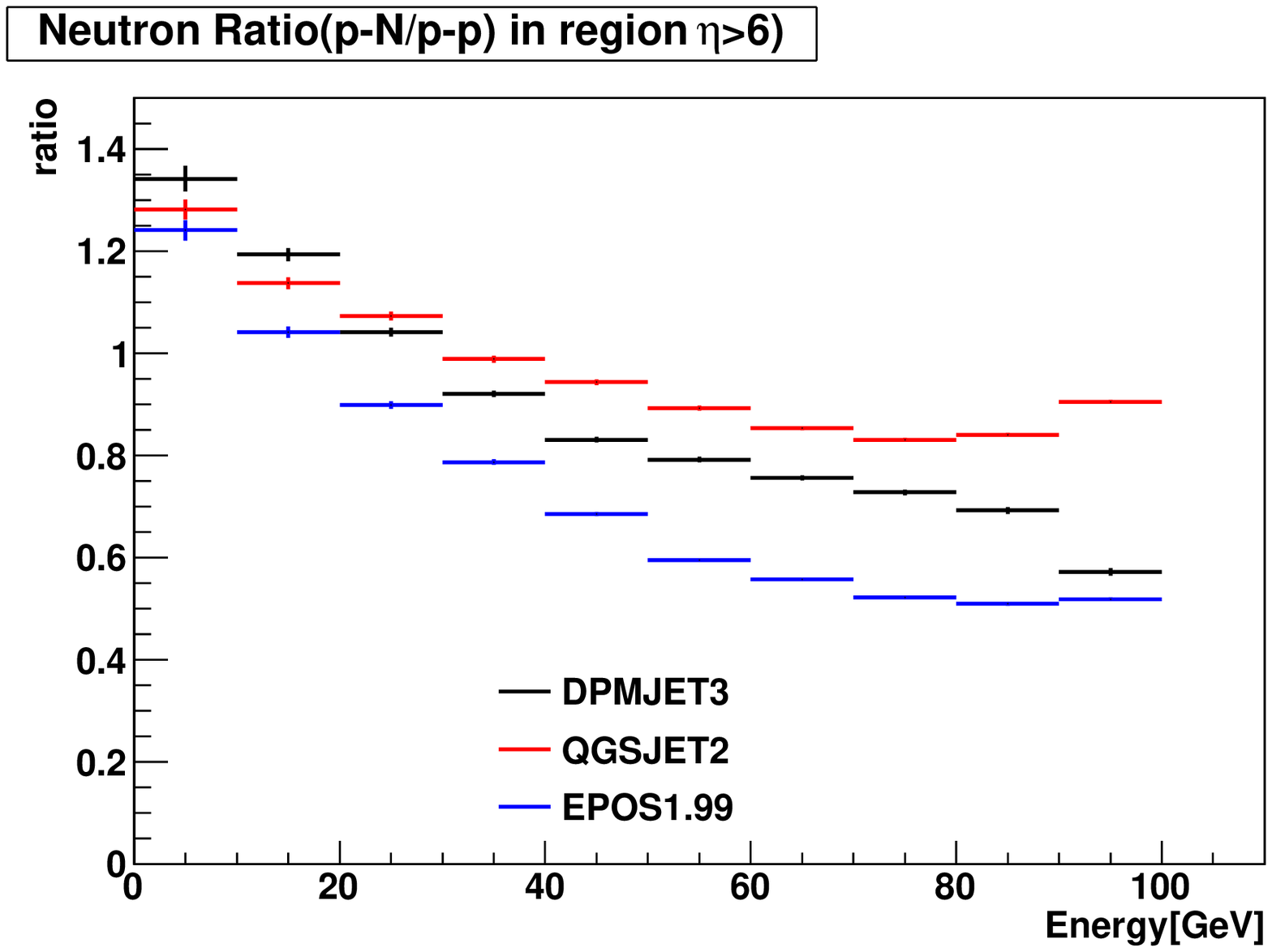}
  \vspace*{8pt}
  \caption{Nuclear effect by different models.  The left figure is for photons, and the right figure is for neutrons.
  Definition in the text.
  \label{fig-nuc-effect}}
  \end{center}
  \end{figure}

  \begin{figure}
  \begin{center}
  \includegraphics[width=7cm]{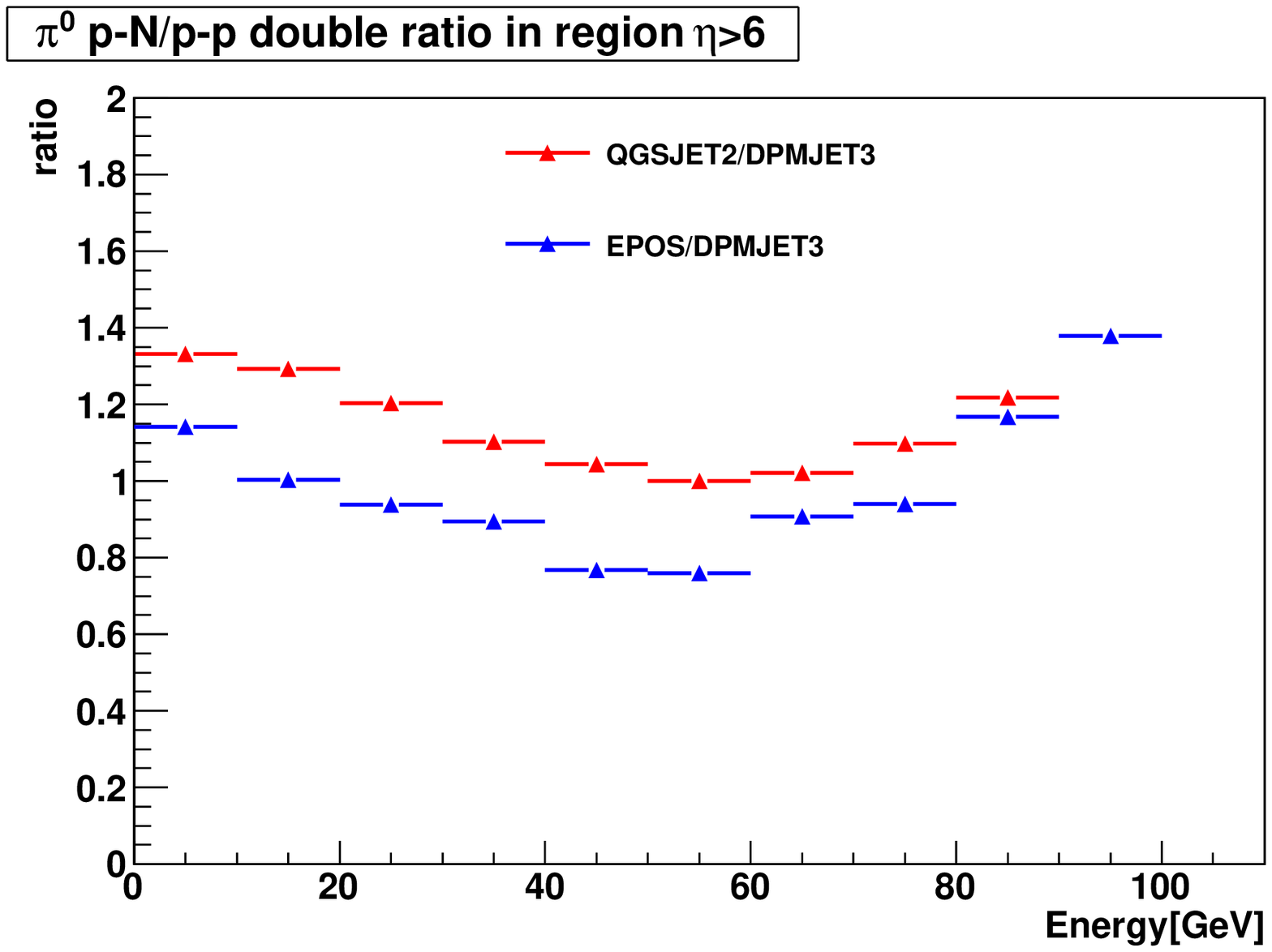}
  \includegraphics[width=7cm]{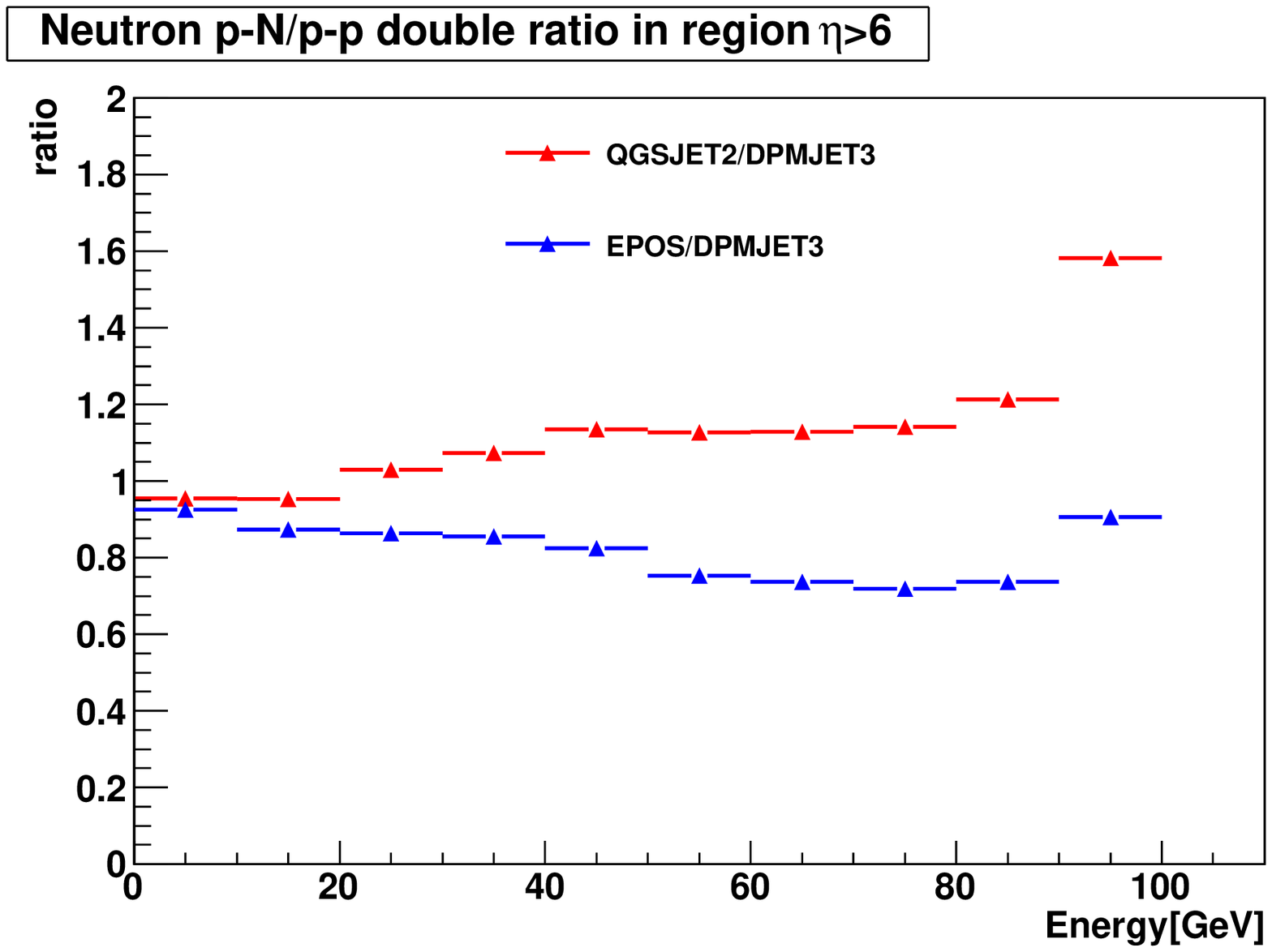}
  \vspace*{8pt}
  \caption{Double ratios of the nuclear effect.  The left figure is for photons and the right figure is for neutrons. 
  \label{fig-nuc-double}}
  \end{center}
  \end{figure}
  
Information from the central detectors and the ZDC indicating the impact parameter (b) of the collision will
help to explore the fundamental process.
As shown in Fig.\ref{fig-b-dep}, the particle spectra produced in peripheral collisions -- here, b$>$2.5\,fm --
are approximately identical to those of p-p collisions.
However, the particles produced in the collisions of b$<$2.5\,fm suffer from multiple collisions
in the nuclei.
Some indicators of the impact parameter are useful to enhance the effect of multiple collisions. 
 
  \begin{figure}
  \begin{center}
  \includegraphics[width=7cm]{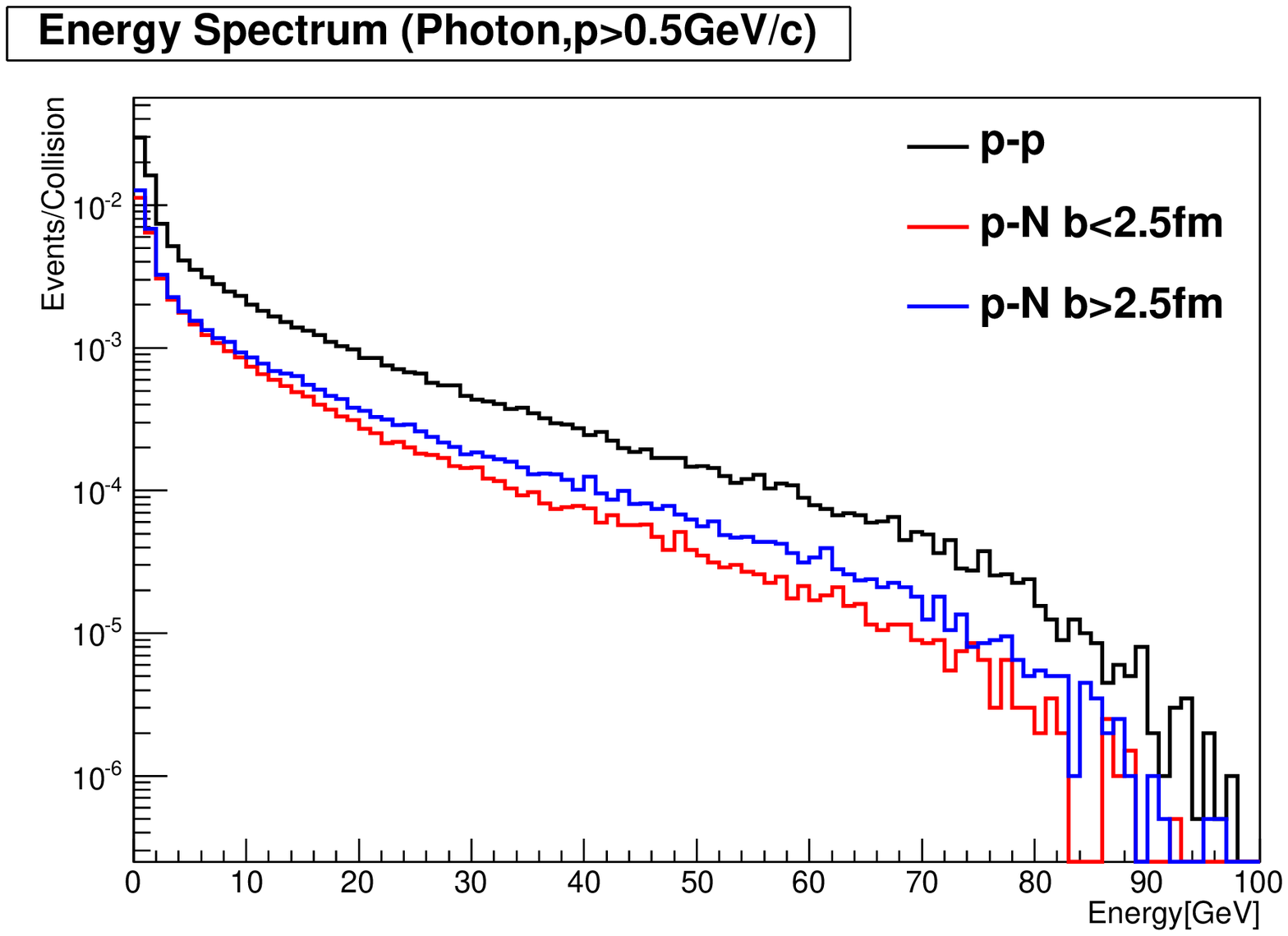}
  \includegraphics[width=7cm]{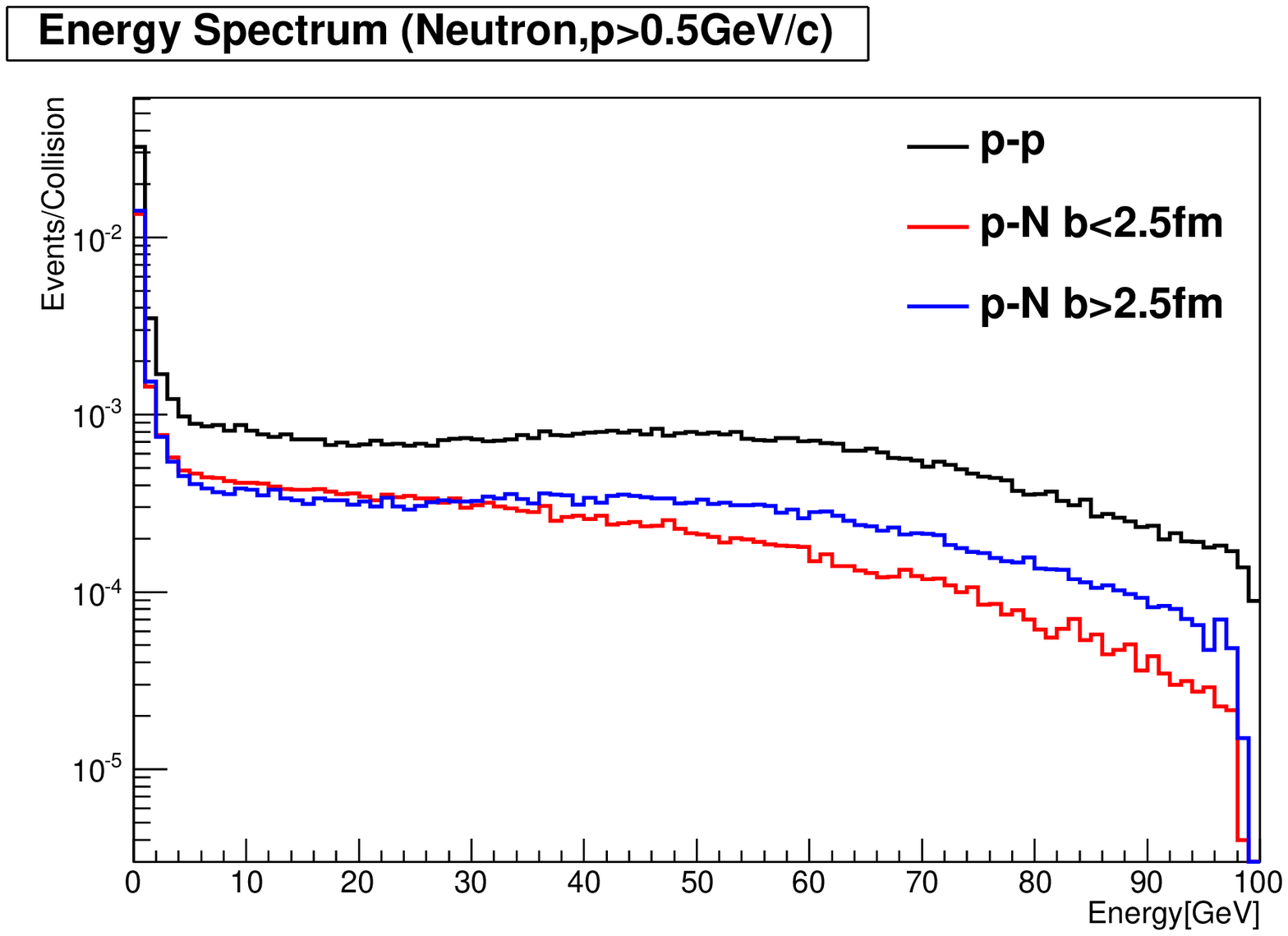}
  \vspace*{8pt}
  \caption{Forward particle spectra for the interaction of b$<$2.5\,fm and b$>$2.5\,fm where b indicates
  the impact parameter.
  \label{fig-b-dep}}
  \end{center}
  \end{figure}
 
To complete the possible cosmic-ray and atmosphere interactions, N-N and Fe-N collisions are also
interesting as future options.
Studying the nuclear effects of light-ion collisions can generally contribute to the physics of cold nuclear matter.
The inclusion of p-Pb collision data at the LHC and possible p-Au collisions at RHIC will provide very wide 
coverage of the nuclear effect in both the mass (A) and energy (E) spaces.

\section{Spin asymmetry measurement} \label{sec-spin}
The origin of the nucleon spin 1/2 has been investigated with polarized p-p collisions at RHIC. 
With the first polarized p-p collisions at $\sqrt{s}$ = 200\,GeV at RHIC, 
a large single transverse-spin asymmetry ($A_N$) for neutron production 
in very forward kinematics was discovered by a polarimeter development experiment \cite{Fukao:2006vd}.
The discovery of the large $A_N$ for neutron production is new, important information to understand the 
production mechanism of the very forward neutron. 
The cross section of very forward neutron production was measured at ISR and Fermilab \cite{Engler:1974nz}. 
These researchers measured a forward peak in the $x_F$ distribution around $x_F$ = 0.8 
and found only a small $\sqrt{s}$ dependence. 
The cross section was also measured at HERA in e-p collisions \cite{Chekanov:2007tv}. 
These researchers found a suppression of the forward peak. 
CERN-NA49 measured the very forward neutron cross section in the p-p reaction \cite{Anticic:2009wd}. 
These researchers' cross section was twice large as those measured at ISR and 
Fermilab, and was consistent with that at HERA. 
To understand the production mechanism of the forward neutron, 
more data are necessary, and the asymmetry measurement gives new information. 

The cross section and $A_N$ of very forward neutron production in polarized p-p collisions was 
measured at PHENIX with a ZDC by adding a position-sensitive SMD (Shower Maximum Detector) 
\cite{Adler:2000bd}. 
The detectors are located downstream of the RHIC-DX dipole magnet so that 
charged particles from collisions are swept out. 
As a beam luminosity monitor, beam beam counters (BBCs) are used. 
These counters are mounted around the beam pipe located $\pm$144\,cm away from the collision point and 
cover $\pm$(3.0--3.9) and 2$\pi$ in the pseudorapidity and azimuth spaces, respectively. 
The data was collected by two sets of triggers for the neutron measurement. 
One trigger was the ZDC trigger for neutron-inclusive measurements by requiring 
energy deposition in either side of the ZDC (the north side or the south side) above 5\,GeV. 
The other trigger was the ZDC$\otimes$BBC trigger, a coincidence trigger of the ZDC trigger with BBC hits,
which were defined as one or more charged particles in both sides of BBCs. 
 
  \begin{figure}[hbp]
  \begin{center}
  \includegraphics[width=12cm]{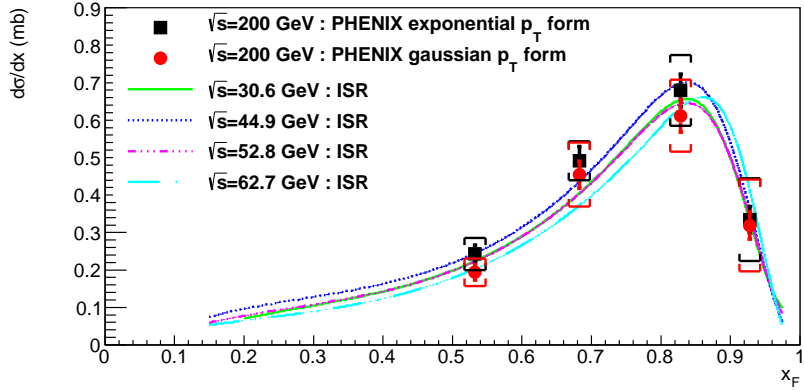}
  \caption{
  The cross section results for forward neutron production in $p+p$ collisions 
  at $\sqrt{s}$=200\,GeV are shown. 
  Statistical uncertainties are shown as error bars for each point, and 
  systematic uncertainties are shown as brackets. 
  Absolute normalization errors for the PHENIX and ISR are 9.7\% and 20\%, 
  respectively. 
  }
  \label{fig:cross-section}
  \end{center}
  \end{figure}

The differential cross section, $d\sigma/dx_F$, in the integrated $p_T$ region, $0 < p_T < 0.11 
\times x_F$\,GeV/$c$, for forward neutron production in p-p collisions at $\sqrt{s}$ = 200\,GeV was 
determined using two $p_T$ distributions: a Gaussian form, as used in the HERA analysis 
\cite{Chekanov:2007tv}, and an exponential form, used for the ISR analysis \cite{Engler:1974nz}. 
The results are plotted in Fig.~\ref{fig:cross-section}. 
The measured cross section was consistent with the ISR result at $\sqrt{s}$ from 30.6 to 62.7 GeV, 
indicating that $x_F$ scaling is satisfied at the higher center-of-mass energy.

  \begin{figure}[htbp]
  \begin{center}
  \includegraphics[width=\hsize]{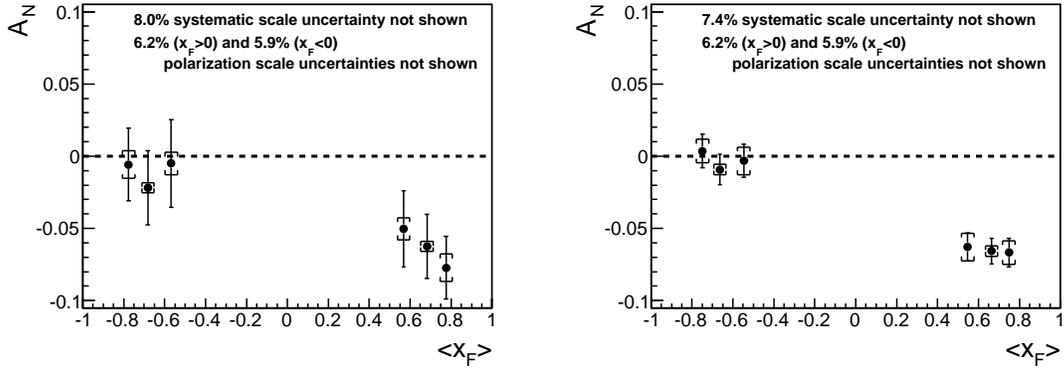}
  \caption{
  The $x_F$ dependence of the $A_N$ for neutron production with the ZDC trigger 
  (left) and with the ZDC$\otimes$BBC trigger (right). 
  The error bars show statistical uncertainties and brackets show 
  $p_T$-correlated systematic uncertainties. 
  }
  \label{fig:asym}
  \end{center}
  \end{figure}

The single transverse-spin asymmetry with the ZDC trigger was $A_N = -0.061 \pm 0.010(stat) 
\pm 0.004(syst)$, and that with the ZDC$\otimes$BBC trigger was $A_N = -0.075 \pm 0.004(stat) \pm 0.004(syst)$. 
The $x_F$ dependence of the $A_N$ of very forward neutron production is shown in Fig.~\ref{fig:asym}.
A significant negative $A_N$ was seen in the positive $x_F$ region and there was no energy 
dependence within the errors in both trigger sets. 
No significant backward neutron asymmetry was observed. 

  \begin{figure}[htbp]
  \begin{center}
  \includegraphics[width=8.5cm]{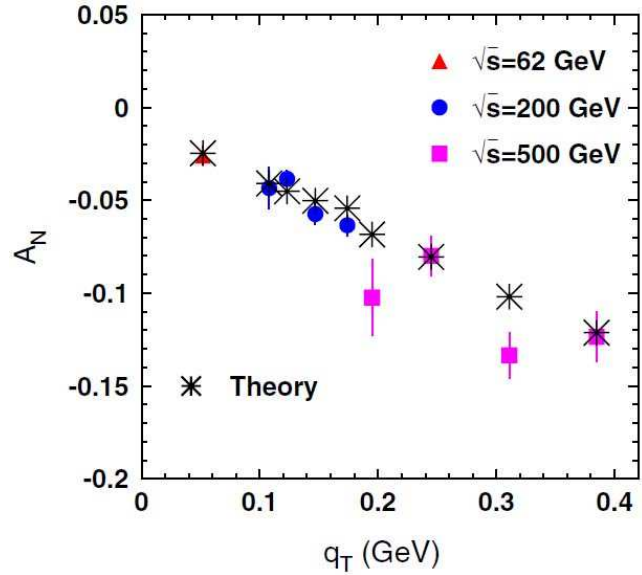}
  \caption{
  The asterisks show the calculation of $A_N$ by Kopeliovich et al. 
  \cite{Kopeliovich:2011bx} compared with measured $A_N$ at $\sqrt{s}$ = 
  62.4\,GeV, 200\,GeV, and\,500 GeV by the PHENIX experiment.
  }
  \label{fig:kopeliovich}
  \end{center}
  \end{figure}

The $\sqrt{s}$ dependence of $A_N$ from three different collision energies, 62.4\,GeV, 200\,GeV, 
and 500\,GeV was also studied. 
The result is shown in Fig.~\ref{fig:asym_pt}. 
The hit position dependence on the detector was measured at each energy, 
although this dependence was largely smeared by the position resolution. 
The result was converted to the $p_T$ dependence,
which showed a hint of the $p_T$ scaling property of the $A_N$ of the very forward neutron production. 
The asymmetry is caused by interference between spin-flip and non-flip amplitudes with a relative phase. 
Kopeliovich et al. \cite{Kopeliovich:2011bx} studied the interference of a pion and $a_1$, or a pion and 
$\rho$ in the $1^+S$ state. 
The data agreed well with the independence of energy, as shown in Fig.~\ref{fig:kopeliovich}. 
The asymmetry is a sensitive to the presence of different mechanisms, 
e.g., Reggeon exchange with spin-non-flip amplitudes even if these amplitudes are small. 

Advantages of the use of the LHCf detector for cross-section and asymmetry measurements at RHIC 
are an improved $p_T$ resolution due to the good position resolution of the LHCf detector and a 
wider $p_T$ coverage due to the detector geometry. 
Improved $p_T$ information enables us to measure the $p_T$ dependence of the cross-section and 
the asymmetry and to determine the invariant cross section, which are useful for the study of the production 
mechanism of forward neutron production as additional and new inputs.

\chapter{Experimental setup and detector} \label{sec-setup}

  \begin{figure}[t]
  \begin{center}
  \includegraphics[width=10cm]{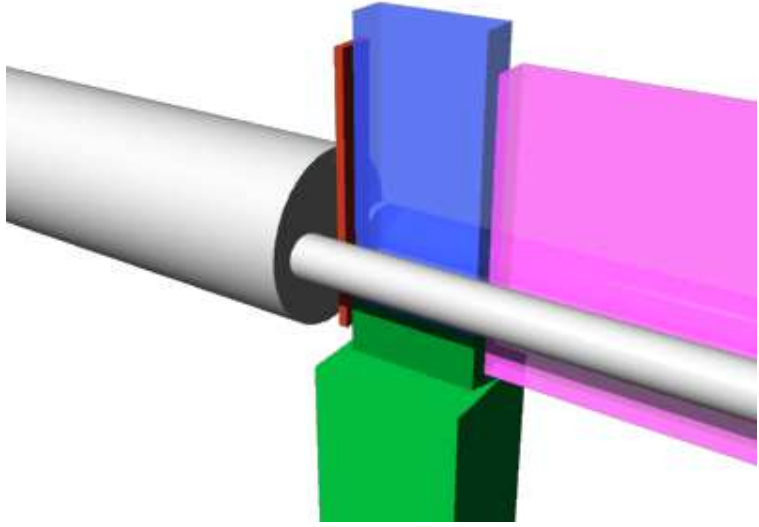}
  \vspace*{8pt}
  \caption{Simple drawing to show the installation location and relationship between beam pipes (grey), 
  ZDC (magenta), RHICf (blue) and front detector (red).
  Green structure indicates the place to install a manipulating lift.
  \label{fig-setup}}
  \end{center}
  \end{figure}

\section{Experimental site and location}
To classify the event category, for example, by impact parameter, the information from the other
detectors will be useful.
From this point of view, installation at the PHENIX or STAR interaction points is preferable.
The detector will be installed at the installation location of the ZDC, 18\,m from the IP, as shown in Fig.\ref{fig-setup}.
The 10\,cm gap between the beam pipes allows an installation of the existing LHCf detector, whose
dimensions  are 92\,mm (W)$\times$290\,mm (L)$\times$620\,mm (H).
The detector will be installed on a mechanical structure, a manipulator, that allows the detector to move vertically.
The installation of the detector in front of the ZDC requires an agreement with the host experiment and the 
accelerator operation team.
As discussed in Chap.\ref{sec-require}, because our operation will be completed within a short period,
the interference to the ZDC measurement for luminosity determination can be minimized.
Using the manipulator, the detector will be placed out of the operation position when RHICf does not take data.
The manipulator also allows the position scan during physics runs to enlarge the p$_{T}$ coverage.

  \begin{figure}
  \begin{center}
  \includegraphics[width=10cm]{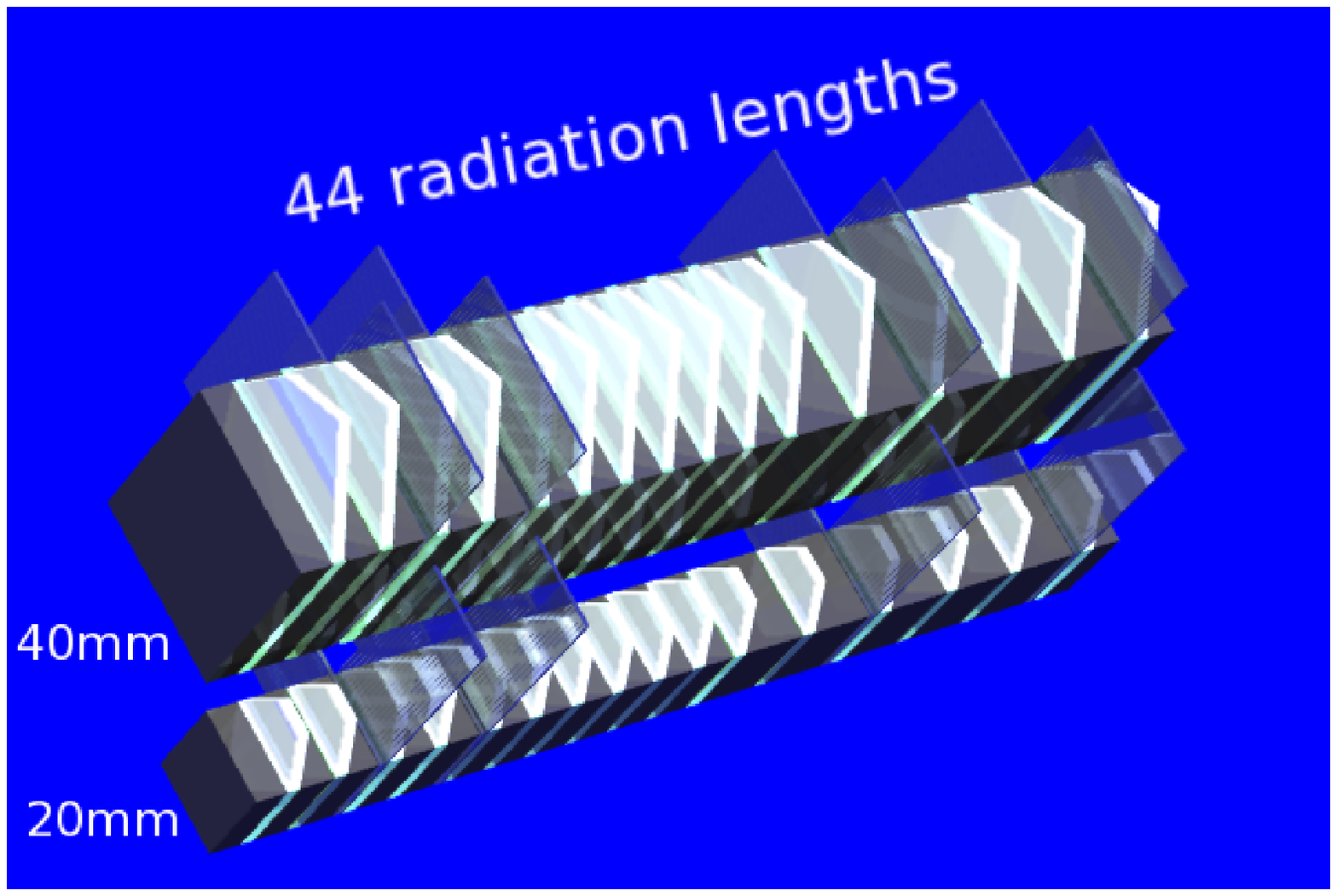}
  \vspace*{8pt}
  \caption{Schematic view of the calorimeters in the LHCf Arm1 detector.
  \label{fig-arm1}}
  \end{center}
  \end{figure}

\section{Detector and data taking}
Here, we suppose to use one of the LHCf detectors, Arm1.
Any upgrade for optimization to the RHIC condition depends on the schedule, budget and manpower as discussed
in Chap.\ref{sec-schedule}.
The Arm1 detector is composed of calorimeters of 20\,mm$\times$20\,mm and 40\,mm$\times$40\,mm
cross sections as shown in Fig.\ref{fig-arm1}.
Each calorimeter has 44 radiation lengths of tungsten in the beam direction.
Sixteen layers of sampling scintillator are inserted every two radiation lengths in the first 11 layers and then every 
four radiation lengths.
Four X-Y pairs of position-sensitive layers composed of scintillating fiber (SciFi, 1\,mm$\times$1\,mm cross section) 
belts are inserted at the depths of 6, 10, 30 and 42 radiation lengths.
Plastic scintillators (EJ-260) and SciFis were used until the 2010 LHC operation but are replaced
with Gd$_{2}$SiO$_{5}$ (GSO) scintillators and GSO bars, respectively, for the LHC 2015 run.
This radiation-hard upgrade assures an operation of a LHCf detector at RHIC even after irradiation at LHC in 2015.

Due to the small transverse dimensions, a large fraction of shower particles leak out from the calorimeter.
However, because this fraction is known to be a function of only the incident position and is independent from the
incident energy, the correct energy can be reconstructed using the information from the position sensitive detector.  
Because the basic performance is determined by the structure and sampling interval, the difference of  
scintillator material is not important for performance.
The performance of the current plastic scintillator detector for a 50--200\,GeV electromagnetic shower was well
studied with beam tests at the SPS North area experimental site in CERN \cite{sps-mase}.  
The expected performance at RHIC is discussed in Sec.\ref{sec-basic}.

The target of our measurement is neutral particles; however, some contamination from charged particles is 
expected.
To eliminate such contamination, a simple position-sensitive scintillator is planned to be installed in front
of the main detector.
Any showering event having a hit in the corresponding position of the front detector is identified as 
contamination from a charged particle.   

\chapter{Sensitivity} \label{sec-sensitivity}

In this Chapter, the expected spectra considering the detector performance and the acceptance are
presented.
The performances of the current LHCf detectors were well studied and published in \cite{sps-mase}.
However, for a more realistic estimation at the RHIC operation, the performance of the upgraded detector
using GSO scintillators is introduced here as possible.
A new MC study is also important to extract the performance below $<$100\,GeV, which is not in the
scope at the LHC operation but important at RHIC.
It is important to note that the responses at this energy range can be directly calibrated
with the beam test at SPS planned for the end of 2014.
A possible improvement in the neutron measurement when combined with the ZDC is also discussed.

\section{Basic performance} \label{sec-basic}
\noindent {\bf Performances in photon measurements}

The energy resolution for the upgraded GSO detector based on the MC simulation is shown in
Fig.\ref{fig-eres}. 
The result shows the resolution of 10--3.5\% for incident photon energies of 10--200\,GeV.
The position resolution for the GSO detector is under investigation.
The beam test for the SciFi detector showed the position resolution of 150--200\,$\mu$m 
for the photon incident of 50--200\,GeV \cite{sps-mizuishi}.

  \begin{figure}
  \begin{center}
  \includegraphics[width=7.2cm]{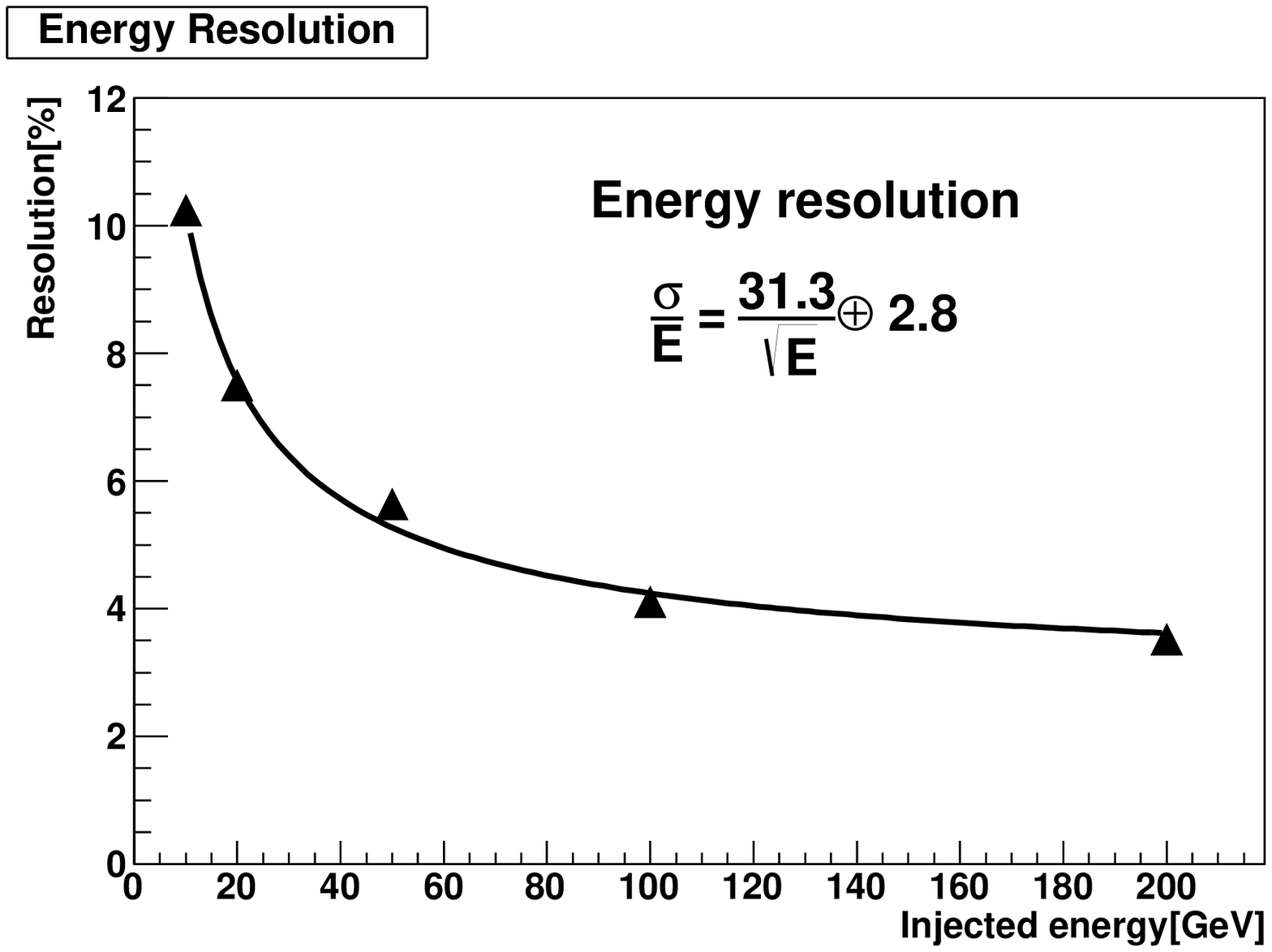}
  \includegraphics[width=7.2cm]{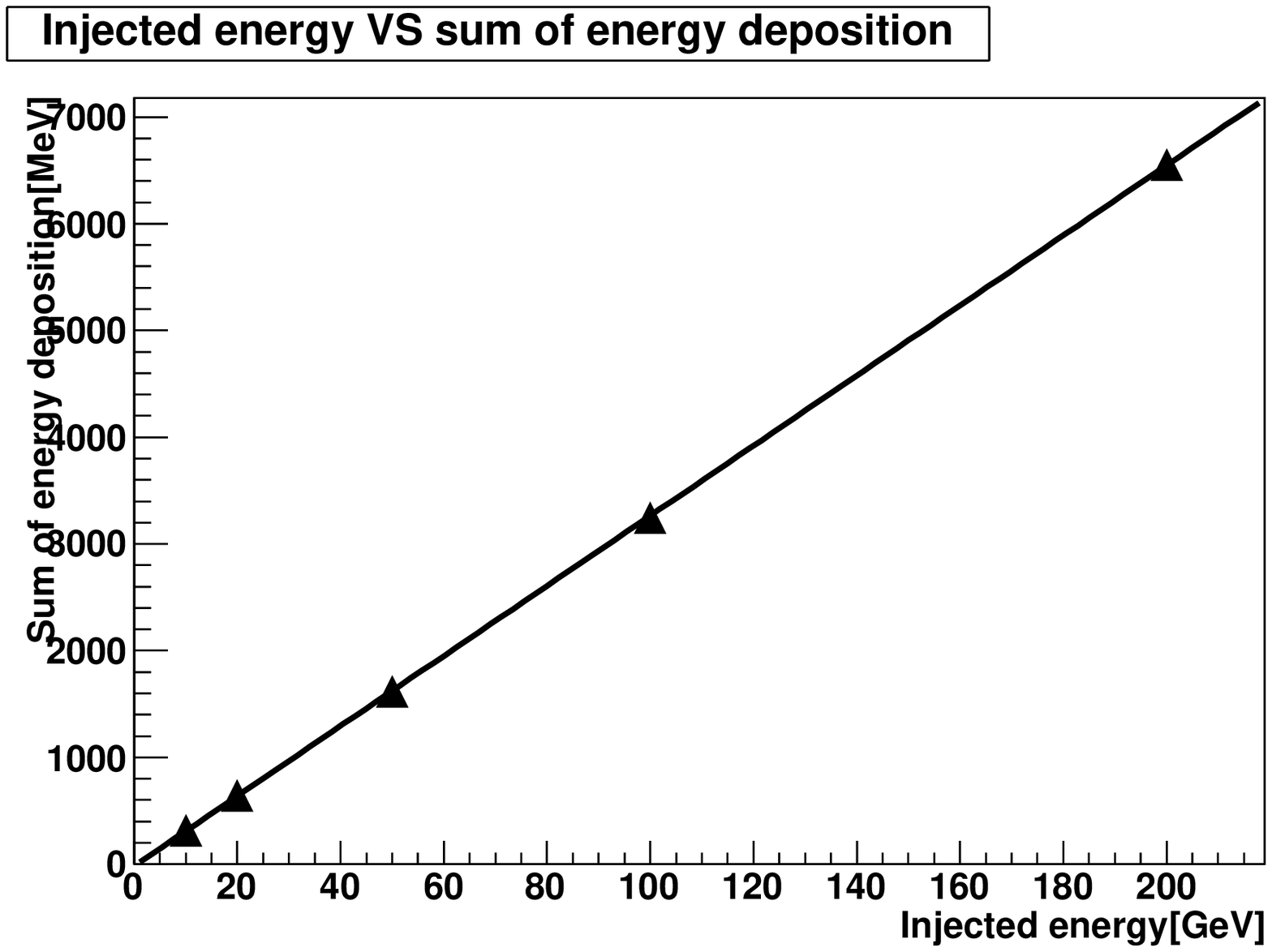}
  \vspace*{8pt}
  \caption{Energy resolution (left) and linearity (right) in photon measurements.
  The results for the GSO detector using MC simulation without electrical noise.
  \label{fig-eres}}
  \end{center}
  \end{figure}

\vskip 5mm \noindent {\bf Performances in neutron measurements}

The performance for incident neutrons is studied only for the plastic scintillator detector.
The energy resolution is 30--35\% with very little energy dependence.
The incident position is determined with a resolution of 2.4--1.8\,mm for incident energies of
100--200\,GeV.

\vskip 5mm \noindent {\bf Combination with ZDC}

A preliminary study for a combined analysis of RHICf and the ZDC indicates an improvement of
the detector performance.
When a neutron-induced hadronic shower develops in the RHICf detector, a better position
resolution than that of the ZDC as described above is obtained.
However, by collecting most of the leaked particles from the small RHICf detector by the
ZDC, an energy resolution close to that of the ZDC (20\% at 100\,GeV) is achieved, as shown in 
Fig.\ref{fig-zdc-combine}.
Taking advantages of two detectors, a better p$_{T}$ resolution for neutrons is expected.
Fig.\ref{fig-ptresolution} demonstrates the expected p$_{T}$ resolution as a function of $x_{F}$ and p$_{T}$
assuming energy resolution of 20\%.
Red and blue bars indicate p$_{T}$ resolutions in cases of the position resolution of 1\,mm and 2.5\,mm, respectively.
The 1\,mm resolution is a typical value in the former SMD measurement introduced in Sec.\ref{sec-spin} while 
2.5\,mm is an expected value in the RHICf experiment.
Significant improvement is expected at low to mid p$_{T}$ range.

  \begin{figure}
  \begin{center}
 \includegraphics[width=9cm]{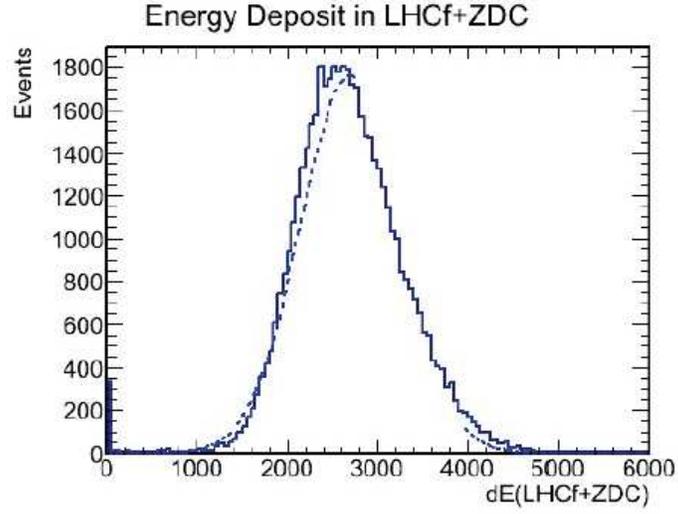}
   \vspace*{8pt}
  \caption{Distribution of deposited energy summed over the RHICf and ZDC detectors for 100\,GeV neutron incident. 
  \label{fig-zdc-combine}}
  \end{center}
  \end{figure}

\  \begin{figure}
  \begin{center}
 \includegraphics[width=9cm]{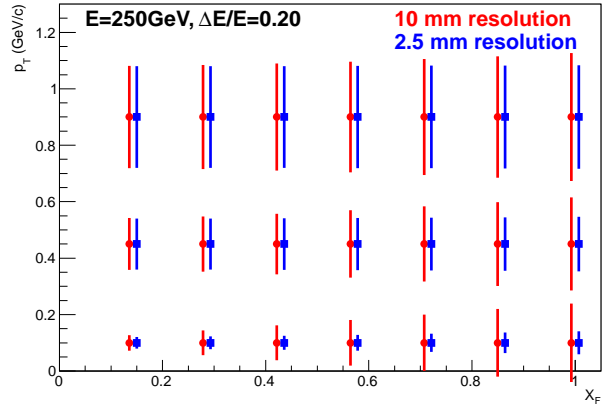}
   \vspace*{8pt}
  \caption{p$_{T}$ resolution as a function of $x_{F}$ and p$_{T}$ in cases of position resolution 10\,mm (red)
  and 2.5\,mm (blue) corresponding to the former and RHICf measurements, respectively.
  \label{fig-ptresolution}}
  \end{center}
  \end{figure}

\section{Sensitivity in 500\,GeV p-p collisions} \label{sec-500gev}
According to the DPMJET3 model, in 3.2\% of inelastic collisions, a neutral particle of $>$10\,GeV is 
observed in any of the two calorimeters.
In this case, the probability of more than one particle hitting a calorimeter is approximately 1\%.
Although such multi-hit events are removed from the analysis, and although this removal causes a systematic 
distortion in the spectral shape, the 1\% multi-hit is at an acceptable level.

As discussed in Chap.\ref{sec-require}, 10$^{6}$ particle detections (3$\times$10$^{7}$ inelastic
collisions at this collision energy) are assumed as a unit of physics operation.
Fig.\ref{fig-single-500} shows the expected spectra for photons and neutrons observed in the RHICf calorimeters.
The energy resolution discussed in Sec.\ref{sec-basic} is taken into account in the solid histograms.
The calculation is performed for the interaction models of DPMJET3 and QGSJET-II.
Even considering the detector response, a statistically clear difference is observed.
In the same data sample, approximately 1,000 of the $\pi^{0}$ are expected to be identified.
The invariant mass spectra of photon pairs are shown in Fig.\ref{fig-pi0-mass}. 
By selecting events with invariant mass of 120--150\,MeV,  the expected $\pi^{0}$ spectra using DPMJET3 and 
QGSJET-II are extracted as shown in Fig.\ref{fig-pi0-500}.
There is a statistically clear difference between the two models.
To provide a differential cross section in $d^{3}\sigma/dp^{3}$, more events are necessary.

  \begin{figure}
  \begin{center}
  \includegraphics[width=7cm]{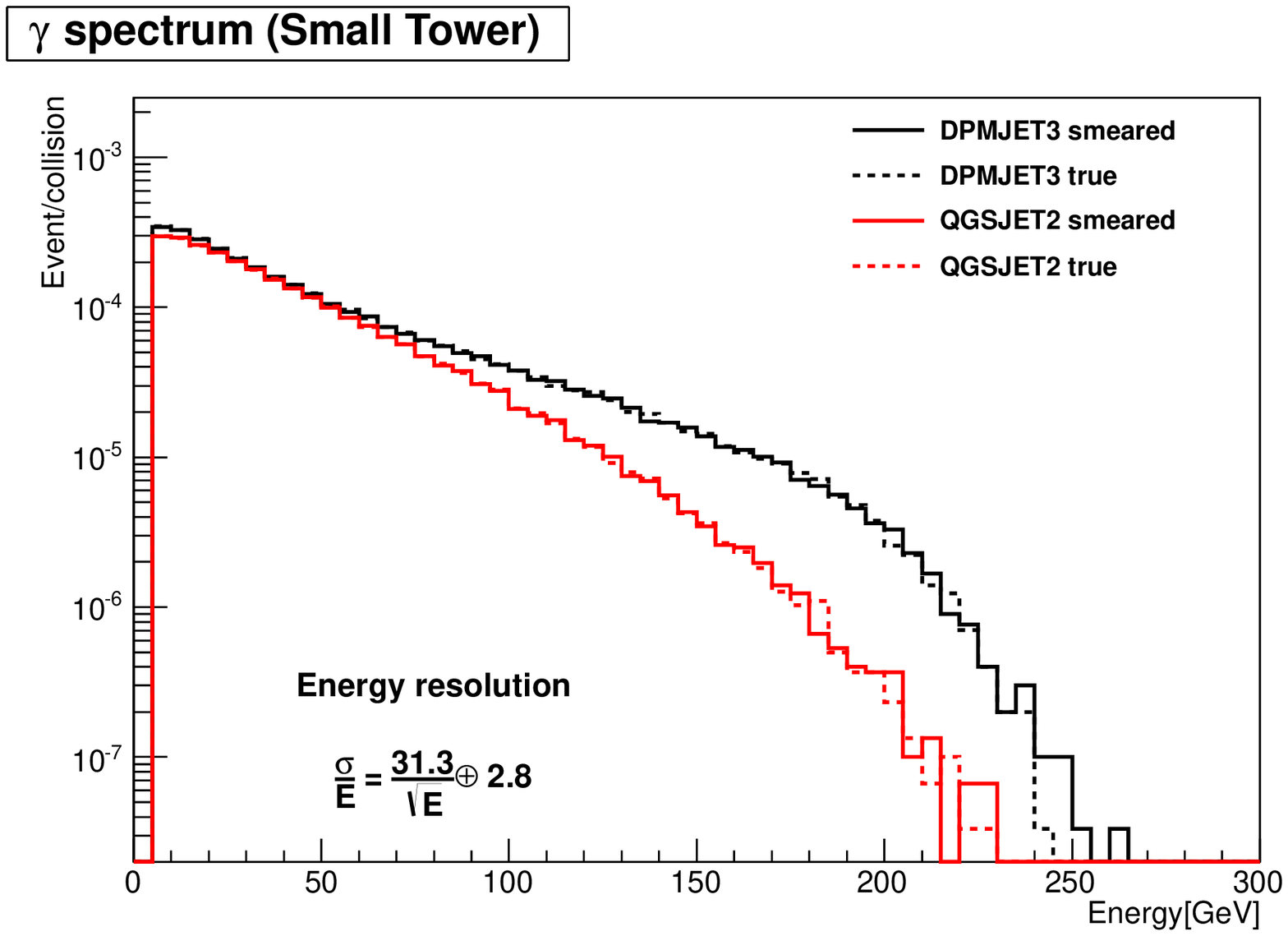}
  \includegraphics[width=7cm]{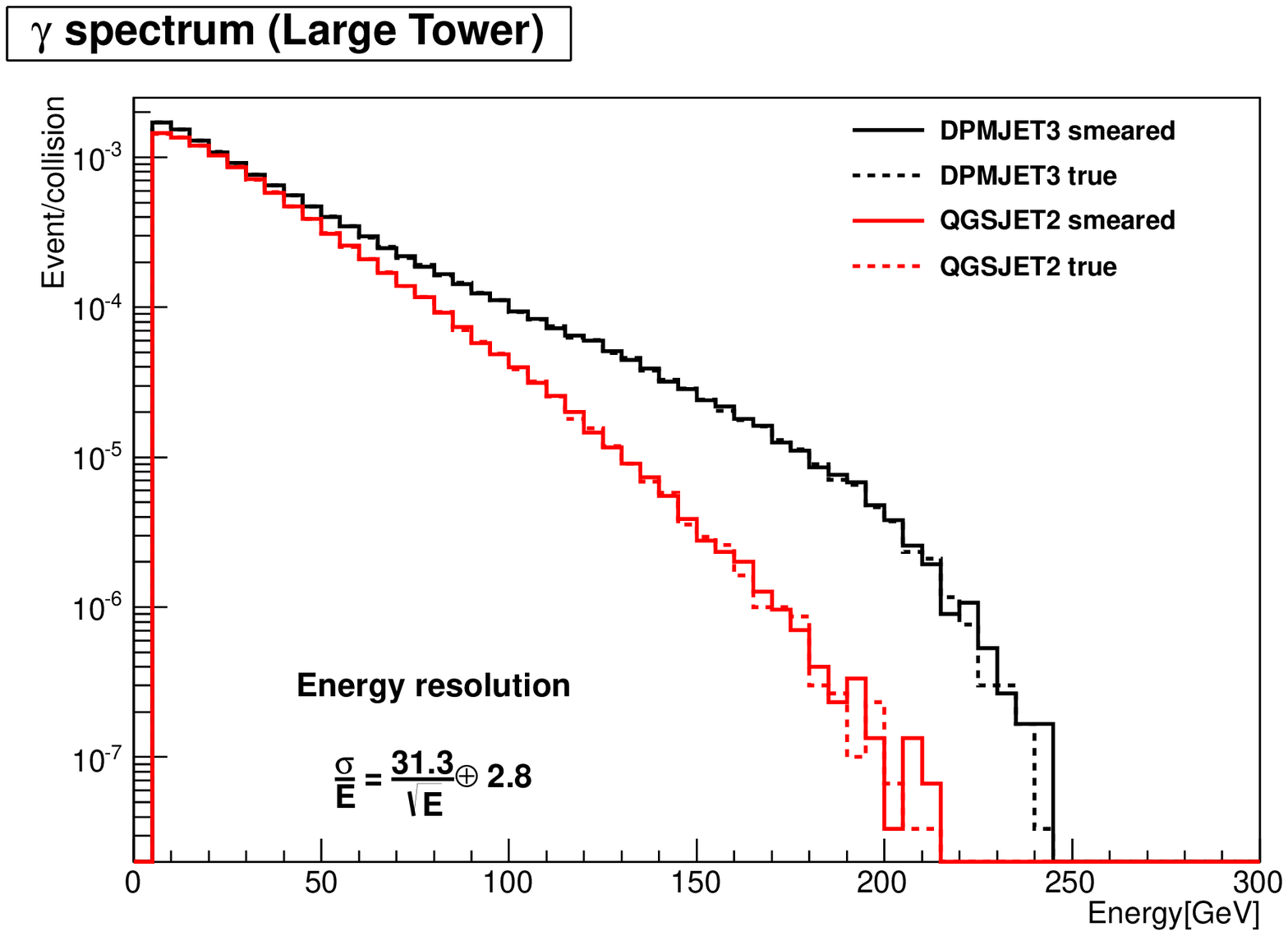}
  \includegraphics[width=7cm]{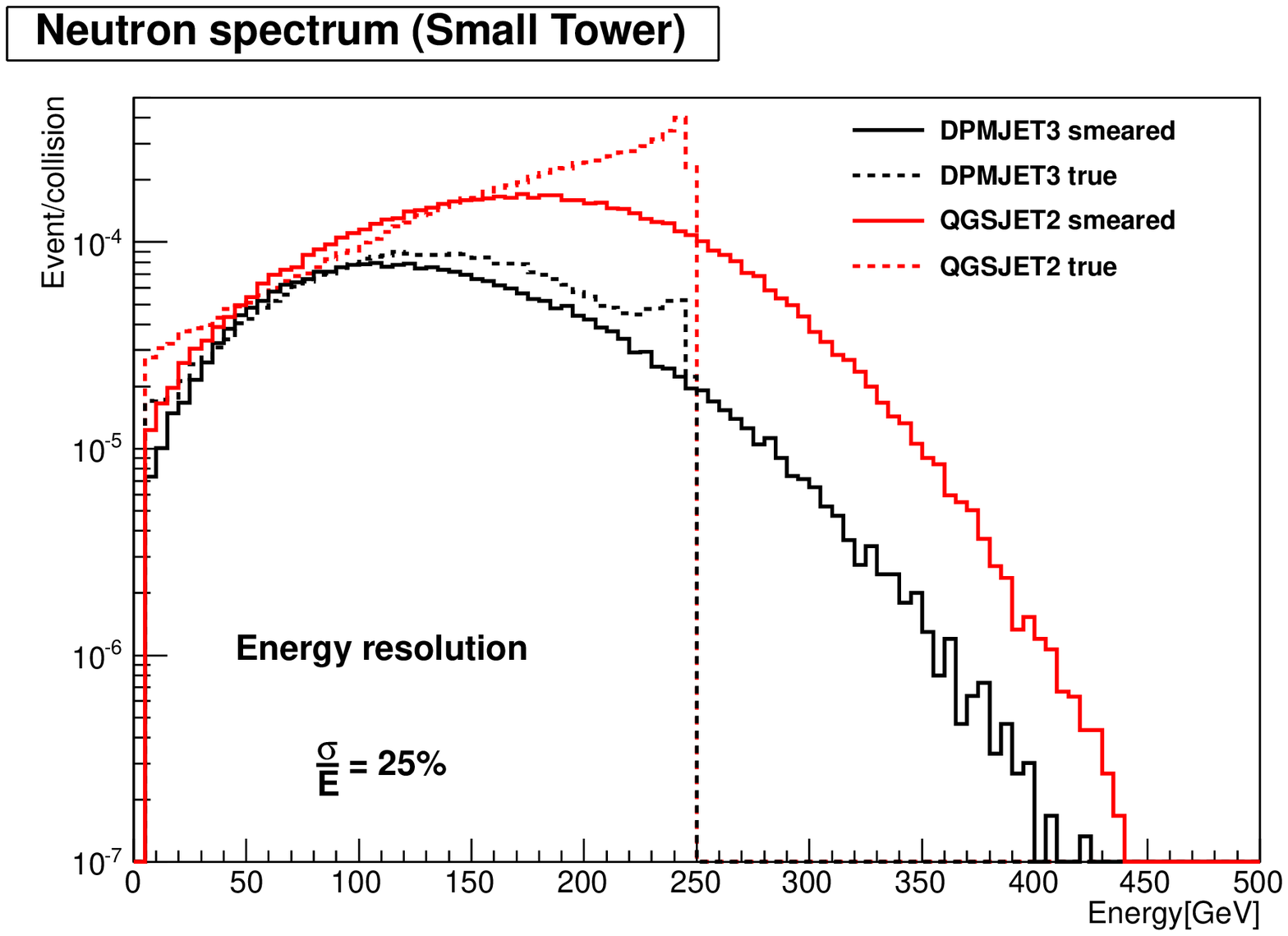}
  \includegraphics[width=7cm]{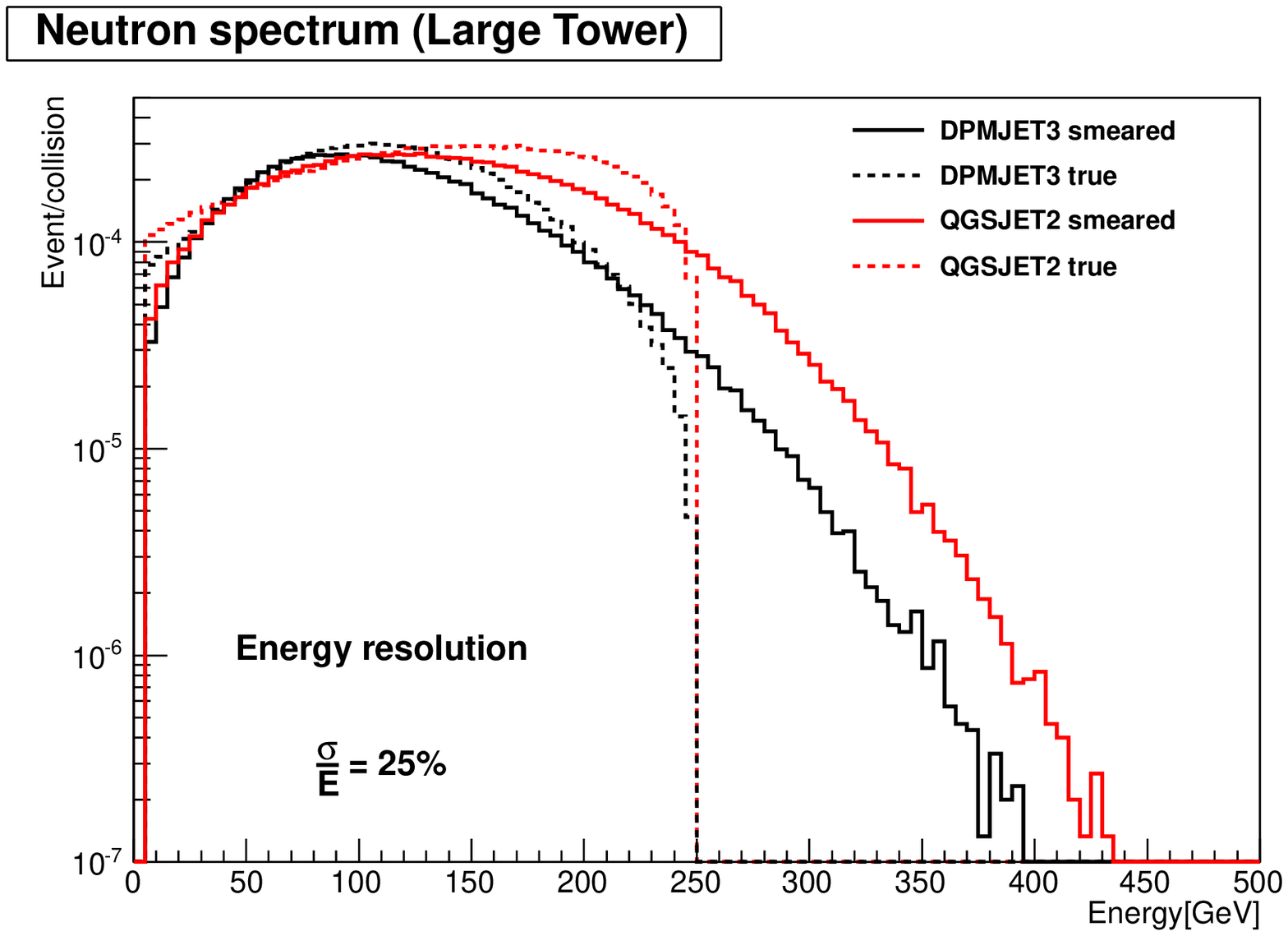}
  \vspace*{8pt}
  \caption{Expected photon (top) and neutron (bottom) spectra obtained by the RHICf calorimeters 
  after 3$\times$10$^{7}$ inelastic interactions.
  Left and right histograms show spectra for the small and large calorimeters, respectively.
  Solid lines are results with energy resolution taken into account while dotted lines without detector response.
  \label{fig-single-500}}
  \end{center}
  \end{figure}
 
  \begin{figure}
  \begin{center}
  \includegraphics[width=10cm]{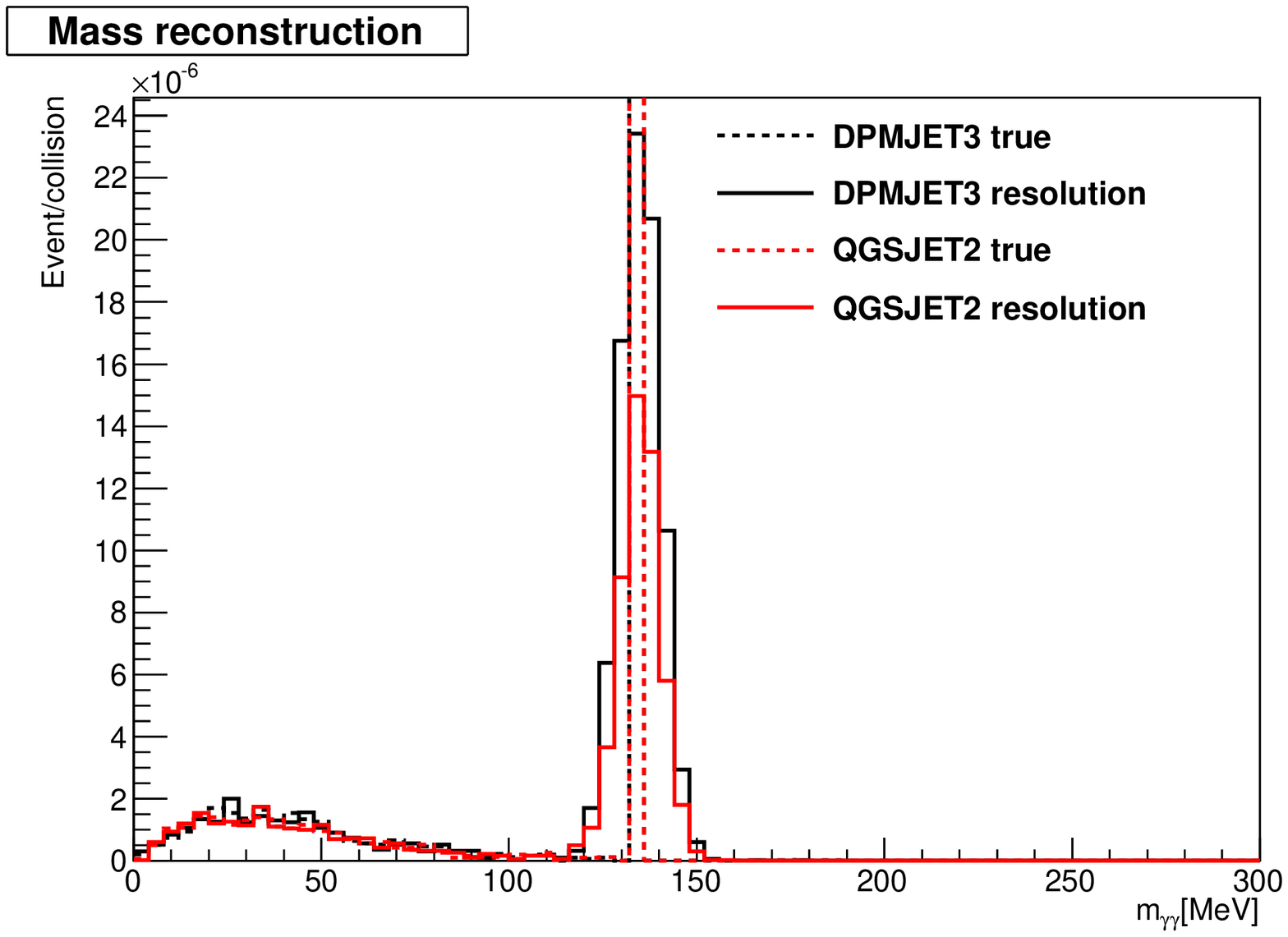}
  \vspace*{8pt}
  \caption{Expected invariant mass spectra obtained by the RHICf detector.
  after 3$\times$10$^{7}$ inelastic interactions.
  Solid and dotted lines for with and without detector response, respectively.
  \label{fig-pi0-mass}}
  \end{center}
  \end{figure}
  
  \begin{figure}
  \begin{center}
  \includegraphics[width=10cm]{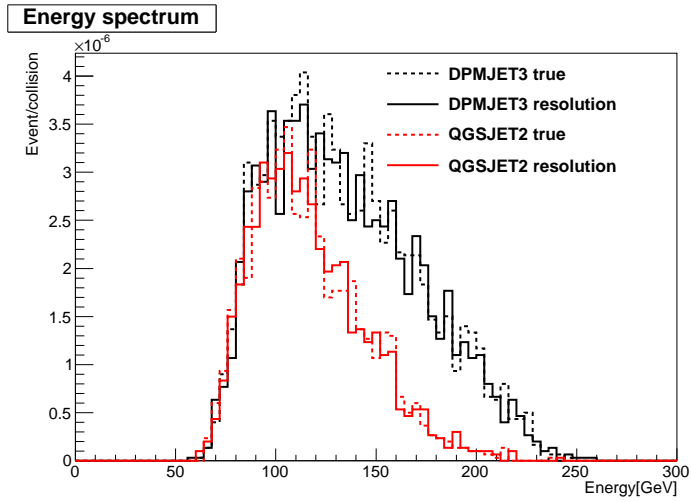}
  \vspace*{8pt}
  \caption{Expected $\pi^{0}$ spectra obtained by the RHICf detector
  after 3$\times$10$^{7}$ inelastic interactions.
  Solid and dotted lines for with and without detector response, respectively.
  \label{fig-pi0-500}}
  \end{center}
  \end{figure}

The expected numbers of events after 0.1\,pb$^{-1}$ of 500\,GeV p-p collisions are studied using the PYTHIA 
minimum-bias event generation.
The numbers of inclusive photons, neutrons, and $\pi^0$ at $x_F > 0.4$ accepted in the LHCf detector
are summarized in Tab.\ref{tbl:statistics}.
The precision of the asymmetry measurement, $\delta A = 1 / P\cdot\sqrt{N}$, is also summarize in the table. 
The polarization $P$ is assumed to be 70\%. 
A 0.1\%-level measurement of neutron asymmetry and a 1\%-level 
measurement of $\pi^{0}$ can be performed up to $p_T$ = 0.7 GeV/$c$. 

  \begin{table}[htbp]
  \begin{center}
  \caption{Statistics (1,000 events) obtained from 0.1 pb$^{-1}$ luminosity.  
  $\delta$A indicates the expected statistical accuracy of the asymmetry determination.}
  \vskip 3mm
  \begin{tabular}{l||r|r|r|r|r|r}
                  & \multicolumn{2}{l|}{neutron} & \multicolumn{2}{l|}{photon}  & \multicolumn{2}{l}{$\pi^0$} \\
  $p_T$ (GeV/$c$) & $N$      & $\delta A$       & $N$      & $\delta A$       & $N$      & $\delta A$ \\
  \hline
  \hline
  0.1 -- 0.2      &  4,790  & 0.0007           &     450 & 0.0021           &   200 & 0.0032 \\
  0.2 -- 0.3      &  7,030  & 0.0005           &  1,220 & 0.0013           &   120 & 0.0041 \\
  0.3 -- 0.4      & 10,290 & 0.0004           &  1,290 & 0.0013           &   160 & 0.0035 \\
  0.4 -- 0.5      &  7,870  & 0.0005           &     600  & 0.0018           &   150 & 0.0037 \\
  0.5 -- 0.6      &  4,520  & 0.0007           &     150  & 0.0037           &    70 & 0.0055 \\
  0.6 -- 0.7      &  1,990  & 0.0010           &       50  & 0.0067           &    20 & 0.0097 \\
  \end{tabular}
  \end{center}
  \label{tbl:statistics} 
  \end{table}

\section{Sensitivity in 200\,GeV p-N collisions}
The expected spectra from the 200\,GeV p-p and p-N collisions are also studied.
The most important difference from the 500\,GeV p-p case is the observation at the N-remnant side.
In this case, in most of the events, the calorimeters record showers from more than one particle
mainly from the fragmentation neutrons.
As discussed in Sec.\ref{sec-500gev}, this high multi-hit rate disables a reliable analysis and
the observation at the N-remnant side is not a target of this proposal.

In the p-p and p-N (p-remnant side) collisions, in 0.05\% collisions, a single particle is recorded
in one of the RHICf calorimeters.
The probability of multi-hit contamination is estimated at below 1\% and at the acceptable level.

The expected spectra of the p-N/p-p ratio introduced in Sec.\ref{sec-phys-nuc} after 2$\times$10$^{7}$
collisions are shown in Fig.\ref{fig-nuc-exp}.
Similar trend to Fig.\ref{fig-nuc-effect} which does not take into account the experimental effects is
clearly visible.
It is also noted that Fig.\ref{fig-nuc-effect} shows the nuclear effect in the $\pi^{0}$ spectrum but 
Fig.\ref{fig-nuc-exp} is for photons.
The difference in the low energy part  is due to the background particles produced in the interaction 
between collision products and beam pipe.

  \begin{figure}
  \begin{center}
  \includegraphics[width=11cm]{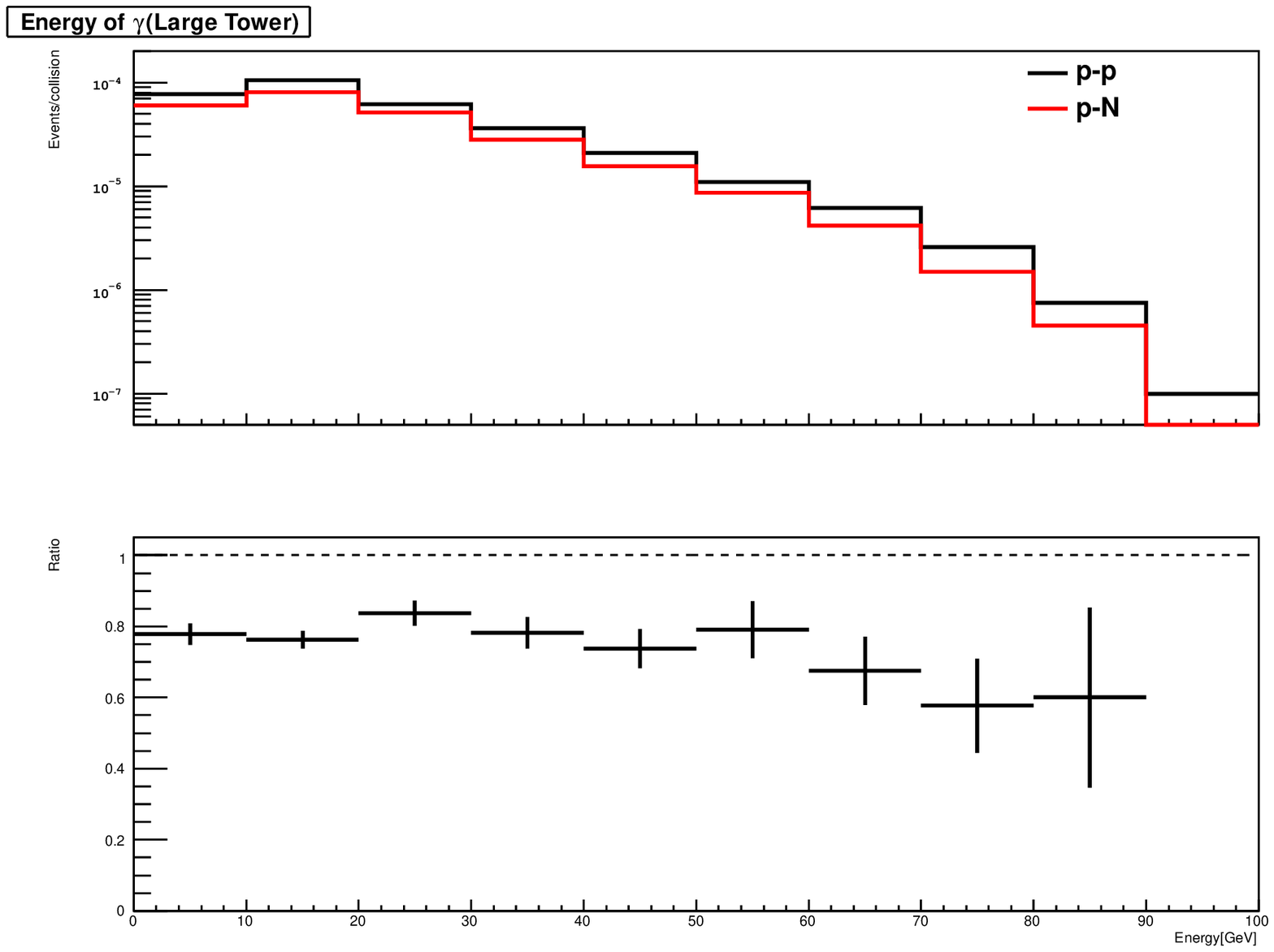}
  \includegraphics[width=11cm]{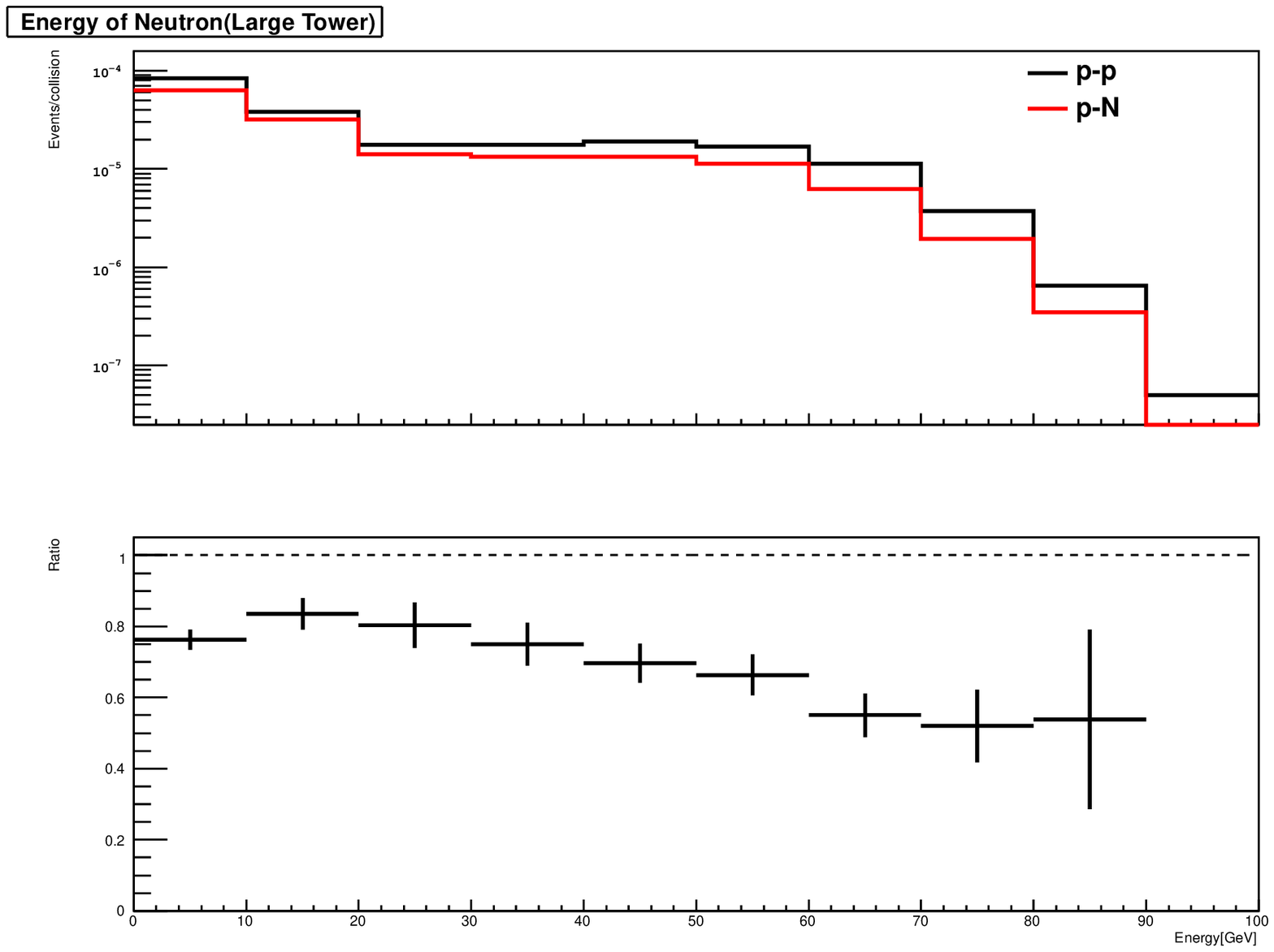}
  \vspace*{8pt}
  \caption{Expected spectra of p-N/p-p ratio obtained by the RHICf large calorimeter
  after 2$\times$10$^{7}$ inelastic interactions.
  Top two plots show the spectra of photons in p-p (black) and p-N (red) collisions and its ratio p-N/p-p.
  Bottom two plots are for neutrons.
  \label{fig-nuc-exp}}
  \end{center}
  \end{figure}

\chapter{Beam condition and requirements} \label{sec-require}
\section{Beam condition and operation time}
Assuming the use of the LHCf Arm1 detector for RHICf, the data acquisition speed is limited to 1\,kHz.
According to the DPMJET3 simulation, the geometrical acceptance of the RHICf detector for $>$10\,GeV 
neutral particles is 3.2\% for an inelastic collision in the case of  500\,GeV p-p collisions.
This situation requires a desirable collision rate of 31\,kHz.
Using a cross section of inelastic collision $\sigma_{ine}$ = 50\,mb, an ideal luminosity is  
6.3$\times$10$^{29}$\,cm$^{-2}$s$^{-1}$.
To avoid a signal overlap in the shaping amplifier of the MAPMT readout electronics, a desirable interval of 
events is $>$ 2\,$\mu$sec.
This situation requires the maximum number of bunches and the luminosity per bunch to be 6 and 
1.1$\times$10$^{29}$\,cm$^{-2}$s$^{-1}$, respectively.
The collision pileup $\mu$=0.065, in this case, does not have a significant effect when the detector acceptance of 
3.2\% is taken into account.
It is noted that the 2\,$\mu$sec limit is only to maximize the efficiency.
It is possible to identify the overlapped events and remove them at the offline level.
A summary of the ideal parameters is given in Tab.\ref{tab-stat} along with the cases of 200\,GeV p-p and 
p-N collisions.

To avoid an angular divergence, higher $\beta^{*}$ is needed.
A 55\,$\mu$rad divergence of the collision angle results a 1\,mm positional divergence on the RHICf detector
located 18\,m from the IP, which is a sufficient value to use the position resolution of the detector effectively. 
With the normalized beam emittance $\epsilon^{*}$, the beam energy E and $\beta^{*}$, the angular beam
divergence $\Delta\theta$ at the IP is expressed as
\[ \Delta\theta = 136\,\mu rad \times \sqrt{\frac{(\epsilon^{*}/20\,mm\,mrad)}{(E/100\,GeV)(\beta^{*}/10\,m)}} \]
Assuming nominal values $\epsilon^{*}$=20\,mm\,mrad and $\beta^{*}$=10\,m, 
$\Delta\theta$ = 136\,$\mu$rad and 86\,$\mu$rad for E=100\,GeV and 250\,GeV, respectively.
In the 250\,GeV case, the dispersion is close to the acceptable value, but in the 100\,GeV case, a smaller 
$\epsilon^{*}$ and a larger $\beta^{*}$ as possible are preferred.
 
With the beam condition above (and Tab.\ref{tab-stat}), 10$^{6}$ events are collected in 1,000\,sec
that can be a minimum set of physics data to obtain the energy spectra of forward photon and neutrons 
at each collision energy and beam type.
In the actual operation, several sets of 10$^{6}$ events (a few hours of operation) are required to perform 
a position scan and to study the detector systematics.
Because of this short operation time and the possible interference with the ZDC as described below, 
our operation will be performed as special runs.

For the asymmetry measurement, an integrated luminosity of 0.1 pb$^{-1}$ at 500\,GeV p-p collisions is required
as discussed in Sec.\ref{sec-500gev}.
Assuming a luminosity of  2.5$\times$10$^{30}$\,cm$^{-2}$ s$^{-1}$, 4$\times$10$^4$\,s, or approximately 
11 hours, are required for data acquisition. 
To compensate for the slow data acquisition speed at a relatively higher luminosity, we will set a higher trigger 
threshold to collect only events $>$100\,GeV.
In the neutron asymmetry study, horizontal polarization of the proton beams is required because a wide
$p_{T}$ survey is available only in the vertical direction and up-down asymmetry is expected.

\section{Other requirements}
In all cases, we place the detector in front of the ZDC.
Before the collision condition is stabilized, the detector will be placed at the garage position where no interference
to the ZDC measurement is expected.
Once the stable collision condition is established, the detector is moved in front of the ZDC using the manipulator
described in Sec.\ref{sec-setup}.
In both the p-p and p-N collision cases, operation of the ZDC is desired.
The data behind the RHICf detector will help to improve the energy resolution of neutron events.
The data in the N-remnant side, where we do not install a detector, will help to categorize the collisions.

The location to install the readout electronics is an issue for discussion.
At the LHC environment, because of the heavy radiation, a large part of the electronics are separated from the 
detector by 200\,m cables.
Depending on the radiation condition at RHIC, the installation of approximately 50 cables (each 100\,m in length)
and space to install readout electronics will be needed.
The space and amount of the cables are not large but require additional discussions and arrangements.

To synchronize the readout electronics with the beam, a beam-induced timing signal and a radio frequency 
clock signal are necessary.
Information of the bunch configuration along with the orbit signal are also useful in case the beam timing signal
is not available.
A flag exchange or a common trigger with the host experiment is also required to allow combined analysis to
classify the events and to improve the energy resolution in the neutron measurement.

  \begin{table}
    \caption{Summary of assumption and requests for beam parameters}
    \label{tab-stat}
    \begin{center}
    \begin{tabular}{lcccc}
      \hline
                                                             &                    & \multicolumn{2}{c}{p-p} & p-N (p side) \\ 
                                                                \cline{3-5}
                                                             & Unit            & 500\,GeV & 200\,GeV    & 200\,GeV \\
                                                                 \hline
      RHICf acceptance ($\xi$)          &                   & 0.032 & 5.9$\times$10$^{-4}$ & 4.5$\times$10$^{-4}$ \\
      \hline
      $\sigma_{ine}$                            & mb             & 50 & 40 & 330 \\
      emittance ($\epsilon$)               & mm mrad  & 20& $<$20& $<$20\\
      $\beta^{*}$                                   & m               &  10& $>$10 & $>$10\\
      luminosity (L)                               & 10$^{30}\,$cm$^{-2}$ s$^{-1}$ & 0.63 & 42 & 6.7 \\
      number of bunches (n$_{b}$)  &                    & 6 & 6 & 6\\
      N$_{ine}$ per bunch crossing ($\mu$) &       & 0.065     &  3.6 & 4.7\\
      RHICf event rate ($\xi\sigma_{ine}L$)   & Hz & 10$^{3}$ & 10$^{3}$ & 10$^{3}$ \\
      RHICf event pileup ($\mu\xi$)   &                    & 2.1$\times$10$^{-3}$ &  2.1$\times$10$^{-3}$ & 2.1$\times$10$^{-3}$\\
      number of RHICf event (N$_{RHICf}$) &        & 10$^{6}$ & 10$^{6}$ & 10$^{6}$ \\
      number of collision (N$_{ine}$) &  10$^{8}$  & 0.31 & 17 & 22 \\
      integrated luminosity                   & nb$^{-1}$  & 0.63 & 42 & 6.7\\
      total time (N$_{ine}$/$\sigma_{ine}$/L) & sec & 10$^{3}$ &10$^{3}$ & 10$^{3}$ \\
      \hline  
    \end{tabular} 
    \end{center}
  \end{table}

\chapter{Schedule and budget} \label{sec-schedule}
Our plan is to bring one of the current LHCf detectors after the LHC 13\,TeV p-p collision operation in
early 2015 and to be ready for the 2016 RHIC run.
Depending on the real schedule of the LHC and the beam species and energy at RHIC, the schedule 
must be flexible.

The current LHCf detector is available even after the 2015 operation at the LHC thanks to the radiation-hard
upgrade using GSO scintillators.
A reuse of one of these detectors for the RHIC operation is agreed upon in the LHCf collaboration.
The main cost for the detector is solved.
However, because the LHCf detectors are optimized for TeV photon detection and use slow readout
electronics, several levels of upgrade or the development of another detector optimized for RHIC are discussed 
as options.
Depending on the available budget, a performance improvement from what is described in this letter can be
expected.

To allow the vertical movement of the detector, a manipulator is necessary.
At the LHC, a manipulator is fixed on the TAN, a neutral particle absorber surrounding the beam pipe around 
the ZDC slot.
Because there is no such structure at RHIC, a dedicated manipulator must be constructed.
Because the detector can be held from the bottom in the RHIC situation, a lift type manipulator is possible and
can be easily constructed with an approximate cost of 10\,kUSD.

\appendix
\chapter{Addendum}
This appendix is added after the submission of LOI.
According to the reference `RHIC Collider Projections (FY 2014 -- FY 2018), W. Fischer et al.', we found some of our
requested luminosity values were higher than the planned values.
We relaxed some of our requirements to be consistent with these values.
The average luminosities at 510\,GeV p-p collisions, 200\,GeV p-p collisions and 200\,GeV p-C collisions are 
documented as
\[ 1.7\times10^{32} \times \Bigl( \frac{\beta^{*}}{0.65\,m} \Bigr)^{-1} \Bigl( \frac{n_{b}}{107} \Bigr) ~~ cm^{-2} s^{-1} \]
\[ 3.8\times10^{31} \times \Bigl( \frac{\beta^{*}}{0.85\,m} \Bigr)^{-1} \Bigl( \frac{n_{b}}{107} \Bigr) ~~ cm^{-2} s^{-1} \]
and
\[ 7\times10^{30} \times \Bigl( \frac{\beta^{*}}{0.8\,m} \Bigr)^{-1} \Bigl( \frac{n_{b}}{111} \Bigr) ~~ cm^{-2} s^{-1} \]
, respectively.
The proportional factors to $\beta^{*}$ and $n_{b}$ are added.
Though our original proposal is to perform proton-nitrogen collisions, proton-carbon collisions planned in 
`RHIC Collider Projections' can give us equivalent physics results and then we assume p-C collisions here.
Using our requested parameters, $\beta^{*}$=10\,m and $n_{b}$=6, possible luminosities are
6.2$\times$10$^{29}$\,cm$^{-2}$ s$^{-1}$, 1.8$\times$10$^{29}$\,cm$^{-2}$ s$^{-1}$ and 
3$\times$10$^{28}$\,cm$^{-2}$ s$^{-1}$ for 500\,GeV p-p collisions, 200\,GeV p-p collisions and 200\,GeV 
p-C collisions, respectively.
The values for 200\,GeV collisions clearly conflict with our ideal luminosities in Tab.\ref{tab-stat}.

To be consistent with the values in `RHIC Collider Projections,' we relaxed the $n_{b}$=6 restriction and consider
its effects.
Assuming $n_{b}$=100 and $\beta^{*}$=10\,m, L=3.0$\times$10$^{30}$\,cm$^{-2}$ s$^{-1}$ is achieved at 200\,GeV
p-p collisions.
Using the inelastic cross section and RHICf acceptance in Tab.\ref{tab-stat}, 71\,Hz of event rate is expected.
To collect 10$^{6}$ events, 3.9\,hours of operation is required.
Note that Fig.\ref{fig-nuc-exp} is produced with 2$\times$10$^{7}$ inelastic collisions corresponding to 10$^{4}$
events.
Consequently, several hours of operation still provide statistically sufficient number of events.

With the new condition above, the average number of interaction per bunch crossing, $\mu$, is 1.5$\times$10$^{-2}$.
In this case, the probability a RHICf event occurs within 2\,$\mu$sec after an event can be estimated as
$1-(1-\mu\xi)^{(2\,\mu s/100\,ns)}$ $\sim$ 20$\mu\xi$ = 1.8$\times$10$^{-4}$.
Here a bunch interval 100\,ns is assumed. 
This means signal overlap in the slow shaping amplifier is negligibly small.
It is worth repeating that such events can be removed at the analysis level.

The DAQ inefficiency is estimated as following.
Because our DAQ speed is limited at 1kHz, successive event within 1\,ms are not recorded and makes inefficiency
in the data taking.
The probability of any event occurring within 1\,ms after an event  can be estimated as
 $1-(1-\mu\xi)^{(1\,ms/100\,ns)}$ $\sim$ 0.085.
 This means the DAQ inefficiency due to the slow readout system is at a 10\% level and does not make significant
 effect. 

The update of Tab.\ref{tab-stat} taken new information above is given in Tab.\ref{tab-stat-new}.
For the 500\,GeV p-p collisions, our ideal luminosity can be achieved with the conditions described in 
`RHIC Collider Projections' and our first proposal. 
However the DAQ inefficiency amounts to 64\%.
The effect of non-negligible inefficiency is included in the time estimate in Tab.\ref{tab-stat-new}.

  \begin{table}
    \caption{Summary of assumption and requests for beam parameters}
    \label{tab-stat-new}
    \begin{center}
    \begin{tabular}{lcccc}
      \hline
                                                             &                    & \multicolumn{2}{c}{p-p} & p-N (p side) \\ 
                                                                \cline{3-5}
                                                             & Unit            & 500\,GeV & 200\,GeV    & 200\,GeV \\
                                                                 \hline
      RHICf acceptance ($\xi$)          &                   & 0.032 & 5.9$\times$10$^{-4}$ & 4.5$\times$10$^{-4}$ \\
      \hline
      $\sigma_{ine}$                            & mb             & 50 & 40 & 330 \\
      emittance ($\epsilon$)               & mm mrad  & 20& $<$20& $<$20\\
      $\beta^{*}$                                   & m               &  10& $>$10 & $>$10\\
      luminosity (L)                               & 10$^{30}\,$cm$^{-2}$ s$^{-1}$ & 0.62 & 3.0 & 0.5 \\
      number of bunches (n$_{b}$)  &                    & 6 & 100 & 100\\
      N$_{ine}$ per bunch crossing ($\mu$) &       & 6.5$\times$10$^{-2} $  &  1.5$\times$10$^{-2}$ & 2.1$\times$10$^{-2}$\\
      RHICf event rate ($\xi\sigma_{ine}L$)   & Hz & 10$^{3}$ & 71 & 74 \\
      RHICf event pileup ($\mu\xi$)   &                    & 2.1$\times$10$^{-3}$ &  8.9$\times$10$^{-6}$ & 9.5$\times$10$^{-6}$\\
      RHICf event overlap (20$\mu\xi$) &                   & --- &  1.8$\times$10$^{-4}$ & 1.9$\times$10$^{-4}$\\
      RHICf DAQ inefficiency ($\eta_{ineff}$)      &                    & 0.64 &  0.085 & 0.090\\
      number of RHICf event (N$_{RHICf}$) &        & 10$^{6}$ & 10$^{6}$ & 10$^{6}$ \\
      number of collision (N$_{ine}$) &  10$^{8}$  & 0.31 & 17 & 22 \\
      integrated luminosity                   & nb$^{-1}$  & 0.63 & 42 & 6.7\\
      total time (N$_{ine}$/$\sigma_{ine}$/L/(1-$\eta_{ineff}$)) & hour & 0.77  & 4.3 & 4.1 \\
      \hline  
    \end{tabular} 
    \end{center}
  \end{table}

\end{large}
\end{document}